\documentclass[prd,showpacs,floatfix
]{revtex4}
\usepackage{graphicx}
\usepackage{amsmath}
\usepackage{amssymb}
\usepackage{times}
\newcommand{\be}{\begin{equation}}
\newcommand{\ee}{\end{equation}\noindent}
\newcommand{\bea}{\begin{eqnarray}}
\newcommand{\eea}{\end{eqnarray}}
\newcommand{\nn}{\nonumber}

\newcommand{\maprightb}[1]{\smash{\mathop{
\hbox to 1cm{\rightarrowfill}}\limits_{#1}}}

\newcommand{\bc}{\begin{center}}
\newcommand{\ec}{\end{center}}

\newcommand{\hsm}{\hspace{-1mm}}
\newcommand{\matTwo}{\left(\begin{array}{rr}}
\newcommand{\matThree}{\left(\begin{array}{rrr}}
\newcommand{\emat}{\end{array}\right )}
\newcommand{\detTwo}{\left|\begin{array}{rr}}
\newcommand{\detThree}{\left|\begin{array}{rrr}}
\newcommand{\edet}{\end{array}\right |}
\newcommand{\Det}{{\rm Det}}

\newcommand{\Nred}{N_{\rm red}}
\newcommand{\ZNc}{Z_{3}}

\def\thline{\noalign{\hrule height 1pt}}

\begin{document}

\title{Wilson Fermion Determinant in Lattice QCD}


\author{Keitaro Nagata}
\affiliation{
{\textit Department of Physics, The University of Tokyo,} \\
{\textit Bunkyo-ku, Tokyo 113-0033 JAPAN }
}
\author{Atsushi Nakamura}
\affiliation{
{\textit Research Institute for Information Science and Education, Hiroshima University,}\\
{\textit Higashi-Hiroshima 739-8527 JAPAN}
}

\date{\today}

\begin{abstract}
We present a formula for reducing the rank of Wilson fermions from 
$4 N_c N_x N_y N_z  N_t$ to $4 N_c N_x N_y N_z$ 
keeping the value of its determinant. We analyse eigenvalues of a 
reduced matrix and coefficients $C_n$ in the fugacity expansion of the fermion 
determinant $\sum_n C_n (\exp(\mu/T))^n$, which play an important role in the
canonical formulation, using lattice QCD configurations on a $4^4$ lattice.
Numerically, $\log |C_n|$ varies as $N_x N_y N_z$, and goes easily 
over the standard numerical range; We give a simple cure for that. 
The phase of $C_n$ correlates with the distribution of the Polyakov loop
in the complex plain. These results lay the groundwork for future 
finite density calculations in lattice QCD.
\end{abstract}


\maketitle

\section{Introduction}

QCD at finite temperature and density has been one of the most attracting 
subjects in physics. Many phenomenological models predict that
the QCD phase diagram is expected to have a very rich structure,
and thoroughgoing analyses of heavy ion data have been made to
show that we are sweeping finite temperature and density
regions. See Ref.~\cite{Andronic09}.

First-principle calculations based on QCD are now highly called. 
If such calculations would be at our hand,
their outcomes are also very valuable for many research fields:
high energy heavy ion collisions, the high density interior of 
neutron stars and the last stages of the star evolution.
Needless to say, the inside of nucleus is also a baryon rich
environment, and lots of contributions to nuclear physics 
could be expected.

Unfortunately, the first principle lattice QCD simulation
suffers from the sign problem.
Nevertheless, there have been many progresses such as the reweighting 
method\cite{FodorKatz},
the imaginary chemical potential\cite{DElia,FP}
and the canonical formulation\cite{MillRed,EnKacKarLa99};
now some light is  shed on the QCD phase diagram.

Most of lattice QCD studies with non-zero density were done 
with the use of staggered fermions.  
It is desirable to study lattice QCD with Wilson fermions
because it is free from the fourth-root problem.
At zero density, thanks to several algorithm developments, 
lattice QCD simulations with Wilson fermions are now possible even 
on the physical quark masses.

In case of the lattice QCD simulations with finite chemical
potential $\mu$, often we must handle the fermion determinant $\det \Delta(\mu)$, directly. 
For example, the reweighting method requires a ratio of two
determinants,
\be
\frac{\det \Delta(\mu')}{\det \Delta(\mu)}.
\ee
The density of state method needs the phase information\cite{Gocksch}.
The canonical formulation needs the  Fourier transformation of 
the fermion determinant 
\begin{align}
\det \Delta_n = \frac{1}{2\pi} \int d\left(\frac{\mu_I}{T}\right)  e^{-i n \mu_I /T} \det \Delta(\mu_I),
\end{align}
with the quark number $n$ and the imaginary chemical potential $\mu_I$.
In these approaches, the heaviest part of the numerical calculations
is the evaluation of the determinant. An efficient way of the determinant 
evaluation is highly desirable. 
It is very useful if we can transform the fermion matrix $\Delta$ into 
a compressed one whose rank is less than the original one, and yet it gives 
the same value of the determinant, since the numerical cost to evaluate a 
determinant is usually proportional to the third power of the matrix rank.

Such a transformation was found for the staggered fermion by Gibbs~\cite{Gibbs} 
and Hasenfratz and Toussaint~\cite{Hasenfratz:1991ax}, and used in finite 
density simulations, e.g.~\cite{Kratochvila:2004wz,Kratochvila:2005mk,deForcrand:2006ec,Kratochvila:2006jx}. 
Their method has also an advantage in the canonical formulation. 
With the reduction method, the fermion determinant is expressed in powers of fugacity,
\be
\det\Delta(\mu) = \sum_n C_n 
\left(
e^{\mu/T}
\right)^n. 
\label{Jul1910eq1}
\ee
If we obtain the coefficients $C_n$, the Fourier transformation in the canonical 
formulation is easily carried out. 

A reduction method for Wilson fermions has not been established yet. It is unfeasible 
to apply the method for staggered fermions in ~\cite{Gibbs,Hasenfratz:1991ax} to Wilson fermions in a naive way 
because of singular parts contained in the Wilson fermion matrix. Expansions based 
on the trace-log formula have been proposed for the Wilson fermion determinant~\cite{Alexandru:2005ix,Alexandru:2007bb,Li:2007bj,Li:2008fm,Meng:2008hj,Danzer:2008xs,Gattringer:2009wi,Danzer:2009sr}. 
An efficient method to calculate exactly the Wilson fermion determinant
is valuable for finite density simulations with Wilson fermions.

The purpose of the present work is to construct a reduction method for 
Wilson fermions. In Ref.~\cite{Borici}, Borici derived a reduction method
that can be applied to Wilson fermions, and tested it using a Schwinger 
model (QED2) with staggered fermions. We develop further the method 
of \cite{Borici} and derive a reduction formula, which rearranges the 
Wilson fermion determinant in powers of fugacity and reduces the numerical cost. 

Similar to the method in ~\cite{Gibbs,Hasenfratz:1991ax}, the Wilson fermion matrix is 
expressed in a time-plane block matrix form. Projection operators contained 
in the Wilson fermion matrix make it possible to transform forward and 
backward hopping parts separately. Owing to the property of the 
projection operators, the Wilson fermion matrix is transformed so that 
the determinant in the time-plane block form can be carried out analytically. 
The determinant of the Wilson fermion is then reduced 
into that of a reduced matrix, whose size is smaller than the original one. 
The problem results in the diagonalization of the reduced matrix 
instead of 
the original matrix. Solving the eigenvalue problem for the reduced matrix, the 
Wilson fermion determinant is expressed in powers of fugacity. 

This paper is organized as follow. 
In the next section, we show the reduction method for the Wilson fermions. 
In section~\ref{sec:numerical}, as an illustration, we perform numerical simulations on a small 
$4^4$ lattice and calculate the Wilson fermion determinant using 
the reduction method. We discuss the properties of the coefficients
of the fugacity expansion. 
The results are not to be regarded as physical, due to the small
lattice size, but lay the groundwork for future realistic calculations.
The final section is devoted to a summary.
In the appendix, we give (1) the detail of the calculation of the determinant of 
a permutation matrix $P$ used in the reduction formula, (2) a simple numerical trick to evaluate
the fugacity expansion coefficients, and (3) a possible alternative
formulation.

\section{Framework}
\label{sec:framework}
\subsection{Structure of Fermion Matrix}
We employ the Wilson fermions defined by 
\bea
\Delta(x,x') = \delta_{x,x'}
 &-& \kappa \sum_{i=1}^{3} \left\{
        (r-\gamma_i) U_i(x) \delta_{x',x+\hat{i}}
      + (r+\gamma_i) U_i^{\dagger}(x') \delta_{x',x-\hat{i}} \right\}
\nonumber
\\
  &-& \kappa \left\{
        e^{+\mu}(r-\gamma_4) U_4(x) \delta_{x',x+\hat{4}}
      + e^{-\mu}(r+\gamma_4) U_4^{\dagger}(x') \delta_{x',x-\hat{4}}
\right\} 
\nn \\
&+& S_{Clover}, \nn \\
S_{Clover} = &-& \delta_{x, x^\prime} C_{SW} \kappa \sum_{\mu \le \nu} \sigma_{\mu\nu} 
F_{\mu\nu}.
\label{Wfermion}
\eea
where $r$, $\kappa$ and $\mu$ are the Wilson term, hopping parameter
and chemical potential, respectively. We include the clover term with 
the coefficient $C_{SW}$. 
For later convenience, we divide the quark matrix into three terms 
according to their time dependence
\begin{align}
\Delta &= B -  2 z^{-1} \kappa r_- V - 2 z \kappa r_+ V^\dagger.
\end{align}
Here $r_\pm = (r \pm \gamma_4)/2$ and $z=e^{-\mu}$, and 
\begin{align}
B(x,x') &\equiv   \delta_{x,x'}
 - \kappa \sum_{i=1}^{3} \left\{
        (r-\gamma_i) U_i(x) \delta_{x',x+\hat{i}}
      + (r+\gamma_i) U_i^{\dagger}(x') \delta_{x',x-\hat{i}} \right\}
\nn \\
&+ S_{Clover}, \\
V(x,x') & \equiv 
 U_4(x) \delta_{x',x+\hat{4}}, \\
V^\dagger(x,x') &\equiv
  U_4^{\dagger}(x') \delta_{x',x-\hat{4}}.
\end{align}
They satisfy $V  V^\dagger = I$. 
Note that $r_\pm$ are projection operators in the case that $r=1$.
In a time-plane block matrix form, $B$ and $V$ are given by 
\bea
&&\begin{array}{ccccccccccccccc}
t'\hsm =\hsm1 & & \cdots & &&&&&&&&&&&t'\hsm =\hsm N_t
\end{array}
\nn \\
B
=
\begin{array}{c}
t=1
\\
t=2
\\
t=3
\\
\cdot
\\
\cdot
\\
\cdot
\\
t=N_t
\end{array}
%
%
&&\left(
\begin{array}{c|c|ccc|c|c}
   B_1 & 0 & 0 & \cdots& & 0 & 0
\\ \hline
   0  &B_2 & 0 &\cdots  &  & 0& 0 
\\ \hline
  0 & 0 & B_3 & \cdots & &  & 
\\ 
 \cdots  &\cdots  &\cdots & \cdots& \cdots &\cdots  &\cdots
\\ 
   &  & &\cdots &   & 0 &0
\\ \hline
    0    & 0 & &  \cdots & 0 &B_{N_t-1}& 0 
\\ \hline
0  & 0 &  &\cdots &0& 0 &B_{N_t} \\
\end{array}
\right). 
\eea

\bea
&&V
=
\nonumber
\\
&&
\hspace{-4mm}
\left(
\begin{array}{c|c|ccc|c}
   0 &  U_4(t\hsm=\hsm1) & 0 & \cdots& &  0 
\\ \hline
   0  & 0 &  U_4 (t\hsm=\hsm2)&\cdots  &  &  0 
\\ \hline
  0 & 0 & 0 & \cdots & & 
\\ 
 \cdots  &\cdots  &\cdots & \cdots& \cdots  &\cdots
\\ 
   &  & &\cdots    & U_4(t \hsm={\hsm} N_t-2) & 0 
\\ \hline
    0    & 0 & &  \cdots & 0 &  U_4(t\hsm={\hsm} {N_t\hsm-\hsm 1}) 
\\ \hline
-U_4(t\hsm=\hsm{N_t}) & 0 &  &\cdots &0&  0 \\
\end{array}
\right). 
\nonumber
\eea
%


\subsection{Reduction Formula for Wilson Fermions}
Now, we derive a reduction formula for the Wilson fermions. 
A starting point is to define a matrix~\cite{Borici},
\begin{align}
P = (c_a r_- + c_b r_+ V z^{-1}),
\end{align}
which is referred to as a permutation matrix~\cite{Borici}.
The parameters $c_a$ and $c_b$ are arbitrary scalar except for zero, and may be set to one. 
We can use these parameters to check the following reduction formula numerically.
Since $r_\pm$ are singular, the matrix $P$ must contain both of them;
otherwise $P$ is singular. It is straightforward to check 
$\det(P)  = (c_a c_b z^{-1})^{N/2} $, where 
$N=4N_c N_x N_y  N_z  N_t$. 
Multiplied by $P$, the quark matrix is transformed into
\begin{align}
\Delta P = (c_a B r_- - 2 c_b \kappa r_+ ) 
+ ( c_b B r_+ - 2 c_a \kappa r_-) V z^{-1} . 
\label{May0910eq1}
\end{align}
In the time-plane block matrix form, the first and second terms of 
Eq.~(\ref{May0910eq1}) are given by 
\begin{align}
(c_a B r_- - 2 c_b \kappa r_+ )
&= \left( \begin{array}{cccc} 
 \alpha_1  & & & \\
 & \alpha_2 & & \\
 & & \ddots & \\
 & & & \alpha_{N_t}
\end{array}\right), 
\label{Jun1910eq1} \\
( c_b B r_+ - 2 c_a \kappa r_-) V z^{-1}
& = \left( \begin{array}{ccccc} 
 0 & \beta_1 z^{-1} & & & \\
 & 0 & \beta_2 z^{-1} & & \\
 & & 0 & \ddots & \\
 & & & \ddots & \beta_{N_t-1} z^{-1} \\
 -\beta_{N_t} z^{-1} & & & &  0
\end{array}\right). 
\label{Jul05eq1}
\end{align}
The block-matrices are given by 
\begin{align}
\alpha_i &= \alpha^{ab, \mu\nu}(\vec{x}, \vec{y}, t_i) \nn \\
         &= c_a B^{ab, \mu\sigma}(\vec{x}, \vec{y}, t_i) \; r_{-}^{\sigma\nu} 
         -2  c_b  \kappa \; r_{+}^{\mu\nu} \delta^{ab} \delta(\vec{x}-\vec{y}), 
\\
\beta_i &= \beta^{ab,\mu\nu} (\vec{x}, \vec{y}, t_i), \nn \\ 
        &= c_b B^{ac,\mu\sigma}(\vec{x}, \vec{y}, t_i)\; r_{+}^{\sigma\nu} 
U_4^{cb}(\vec{y}, t_i) -2 c_a \kappa \; r_{-}^{\mu\nu} \delta(\vec{x}-\vec{y}) 
U_4^{ab}(\vec{y}, t_i), 
\end{align}
where the dimensions of $\alpha_i$ and $\beta_i$ are given by 
$N_{\rm red} = N/N_t = 4  N_x N_y  N_z  N_c$. 
We factor out a negative sign caused by anti-periodic boundary conditions 
from the definition of $\beta_{N_t}$. Therefore the negative sign appears 
at the lower-left corner in Eq.~(\ref{Jul05eq1}). The two block-matrices have 
different meaning; 
$\alpha_i$ contains only spatial hopping terms with a fixed time $t=t_i$, 
while $\beta_i$ contains temporal hopping terms as well as spatial ones
due to temporal link variables. 

Combining Eqs.~(\ref{Jun1910eq1}) and (\ref{Jul05eq1}), we can 
carry out the determinant in the time-plane block matrix  
\begin{align}
\det \Delta P & = \left( \begin{array}{ccccc} 
 \alpha_1 & \beta_1 z^{-1} & & &  \\
 & \alpha_2 & \beta_2 z^{-1} & &  \\
 & & \alpha_3 & \ddots &  \\
 & & & \ddots & \beta_{N_t-1} z^{-1}\\
- \beta_{N_t} z^{-1} & & & &  \alpha_{N_t}
\end{array}\right) \nn \\
& = \left(\prod_{i = 1}^{N_t} \det(\alpha_i ) \right) 
\det\left( 1 + z^{-N_t} Q \right) ,
\end{align}
where $Q = (\alpha_1^{-1} \beta_1) \cdots (\alpha_{N_t}^{-1} \beta_{N_t})$, which we refer to as 
a reduced matrix. 
Substituting $\det(P)  = (c_a c_b z^{-1})^{N/2} $, we obtain 
\begin{align}
\det \Delta & = (c_a c_b )^{-N/2} z^{-N/2}  \left(\prod_{i = 1}^{N_t} \det(\alpha_i ) \right) 
\det\left( z^{N_t} +  Q \right).  
\label{May1010eq2}
\end{align}
Here, the rank of the matrices $\alpha_i$ and $Q$ is given by $\Nred = N/N_t$, 
while that of the Wilson fermion is originally given by $N$. 
Thus the reduction formula makes the computation of the determinant $1/N_t^3$ less time. 
Furthermore, the $\mu$ dependent parts are separated from the hopping terms, 
and appear at the overall factor and the second determinant. 

\begin{figure}[htbp]
\begin{center}
\includegraphics[width=0.75\linewidth]{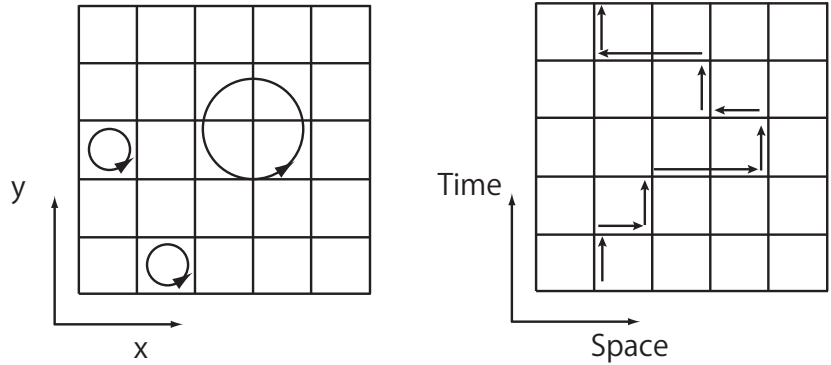}
\begin{minipage}{0.75\linewidth}
\caption{The left panel depicts closed loops on a plane $t=t_i$,
which contribute to $\det \alpha_i$, 
while the right one does paths of quarks from $t=t_1$ to $t=t_{N_t}$, 
which contribute to the reduced matrix $Q$.
}\label{CanJun3010fig1}
\end{minipage}
\end{center}
\end{figure}
\begin{figure}[htbp]
\begin{center}
\includegraphics[width=0.60\linewidth]{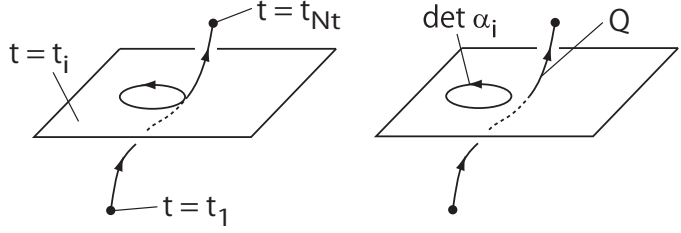}
\begin{minipage}{0.75\linewidth}
\caption{Schematic figures for the reduction procedure.
}\label{CanJul0510fig1}
\end{minipage}
\end{center}
\end{figure}
Equation~(\ref{May1010eq2}) consists of sub-determinants: 
$\prod \det(\alpha_i )$ and $\det( z^{N_t} + Q)$.  
As we have explained, $\alpha_i$ describes spatial hopping terms at a 
time-slice with $t_i$. Hence, $\det \alpha_i$ describes closed loops in a plane 
with a fixed time $t=t_i$ (left panel in Fig.~\ref{CanJun3010fig1}). 
On the other hand, $\beta_i$ contains temporal hopping terms from $t=t_i$ to $t=t_{i+1}$. 
The reduced matrix $Q$ describes paths of the propagation of quarks from $t=t_1$ to 
$t=t_{N_t}$ (right panel in Fig.~\ref{CanJun3010fig1}). 
Thus, the reduction formula separates closed loops at time-slices 
from paths of the propagation of quarks (Fig.~\ref{CanJul0510fig1}). 

Now, we solve an eigenvalue problem $\det ( Q - \lambda I) = 0$.
With the eigenvalues  $\lambda $, the determinant of the reduced matrix is written as
\be
\det(z^{N_t} + Q) = \prod_{n=1}^{N_{\rm red}} (\lambda_n + z^{N_t} ), 
\ee
which is expanded in powers of $z^{N_t}$,  
\begin{align}
z^{-N/2} \prod_{n=1}^{N_{\rm red}} (\lambda_n + z^{N_t} )
& = \sum_{n=-\Nred /2}^{\Nred /2 } c_n (z^{N_t})^n \nn \\
& = c_{-\Nred/2} (z^{N_t})^{\Nred/2}  +\cdots + c_0 + \cdots 
+ c_{\Nred/2} (z^{N_t})^{-\Nred/2},
\label{Eq:FugExpansion}
\end{align}
where we replace $c_n$ by $c_{-n}$ to obtain the second line from the 
first one.
This is an expansion with regard to (inverse) fugacity $z^{N_t} = \exp ( - \mu /T)$.
Equivalently, this can be interpreted as a winding number expansion, because 
$z^{N_t}$ comes from  closed loops that make a round the lattice in the time-direction. 
Note that the expansion Eq.~(\ref{Eq:FugExpansion}) is exactly done and does not involve any approximation.

Finally, we obtain the reduced quark determinant  
\begin{align}
\det \Delta(\mu) &  =  \sum_{n=-\Nred/2}^{\Nred/2} C_n (e^{\mu/T})^n,  
\label{Jun1410eq1}
\end{align}
Here, $C_n = C c_n$ with 
$C = (c_a c_b )^{-N/2}\left(\prod_{i = 1}^{N_t} \det(\alpha_i ) \right)$. 

The coefficients $c_n$ have two properties. 
If a chemical potential is pure imaginary $\mu = i\mu_I$, then $(z)^* = z^{-1}$
and $(\det \Delta(\mu_I) )^* = (\det(\Delta(\mu_I)))$.
These conditions bring about the first property $c_n^* = c_{-n}$. 
Note that $c_{-\Nred/2}=c_{\Nred/2}=\prod_{n=1}^{\Nred} \lambda_n = 1$.

The second property is concerned with the center transformation $\ZNc$. 
Under $\ZNc$ transformation, the time components of the link variables are 
transformed as  
\begin{align}
U_4(t_i) \to w U_4(t_i),
\end{align}
where $ w = \exp(2\pi i /3)$ is an element of $\ZNc$. 
Regarding the $n$-th term in Eq.~(\ref{Jun1410eq1}), 
if the winding number $n$ is a multiple of $N_c$, 
the coefficient $c_n$ is $\ZNc$ invariant, otherwise $c_n$ is not $\ZNc$ invariant. 
Thus, $c_n$ are classified in terms of $\ZNc$
\begin{align}
c_n \cdots \left\{ \begin{array}{ll}
 \mbox{center invariant} & (n = 3 m)\\
 \mbox{center variant}   & (n = 3 m+1, 3m+2 )
\end{array}\right.   
\end{align}
where $m$ is an integer. 
It is known that the center symmetry is explicitly broken in the presence of quarks.
In the quark determinant, the explicit breaking of the center symmetry is caused by 
the terms having winding numbers not multiple of $N_c$.

\section{Numerical Results}
\label{sec:numerical}

In this section, we demonstrate the calculation of the quark 
determinant $\det \Delta(\mu)$ using the reduction formula. 
In order to see the temperature dependence of $\det \Delta(\mu)$, we set 
$\beta=1.85$ and $2.0$. We employ $(\kappa, C_{SW}) = (0.14007, 1.5759) $ 
and  $(0.1369, 1.5058)$ for $\beta = 1.85$ and $2.0$, respectively. 
We perform hybrid Monte Carlo (HMC) simulations  on the $4^4$ lattice 
with 1,000 quench updates and 100 full QCD HMC trajectories as thermalization. 
After the thermalization, we measure the quark determinant 
on four configurations separated by 20 HMC 
trajectories between measurements. Fundamental numerical data of 
the configurations used for the measurements are shown in 
Table~\ref{Jul1810tab1} and \ref{Jul1810tab2}.

\begin{table}[htbp]
\begin{minipage}{0.75\linewidth}
\caption{The values of the determinant with $\mu=0$ , Polyakov loop and plaquette 
for $\beta =1.85$. (i), (ii), (iii) and (iv) correspond to the configurations
measured.
} \label{Jul1810tab1}
\vspace{0.5cm}
\end{minipage}
\begin{tabular}{llcl}
\thline
 & $\det \Delta(0)$ & Polyakov loop & Plaquette \\
\hline 
(i)   &  $3.0957 \times 10^{-19}$  & $ 0.04377 - 0.25418 i$   & $0.53338$ \\
(ii)  &  $2.0921 \times 10^{-21}$  & $-0.03234 + 0.08711 i$   & $0.50668$ \\
(iii) &  $2.2560 \times 10^{-21}$  & $-0.16365 - 0.10135 i$   & $0.52471$ \\
(iv)  &  $5.1115 \times 10^{-18}$  & $ 0.49234 - 0.12163 i$   & $0.53313$ \\
\thline
\end{tabular}
\end{table}
\begin{table}[htbp]
\begin{minipage}{0.75\linewidth}
\caption{The values of the determinant with $\mu=0$, Polyakov loop and plaquette 
for $\beta =2.0$.
}\label{Jul1810tab2}
\vspace{0.5cm}
\end{minipage}
\begin{tabular}{llcl}
\thline
 & $\det \Delta(0)$ & Polyakov loop & Plaquette \\
\hline 
(i)   & $8.7586\times 10^{-12}$   & $  0.37590 +0.0041 i$ & $0.57810$ \\
(ii)  & $3.0329\times 10^{-12}$   & $  0.13827 -0.1978 i$ & $0.57107$ \\
(iii) & $1.1159\times 10^{-12}$   & $ -0.22324 -0.4285 i$ & $0.57491$ \\
(iv)  & $1.2578\times 10^{-12}$   & $ -0.35711 -0.6028 i$ & $0.57954$ \\
\thline
\end{tabular}
\end{table}

\begin{figure}[htbp]
\begin{center}
\includegraphics[width=0.45\linewidth]{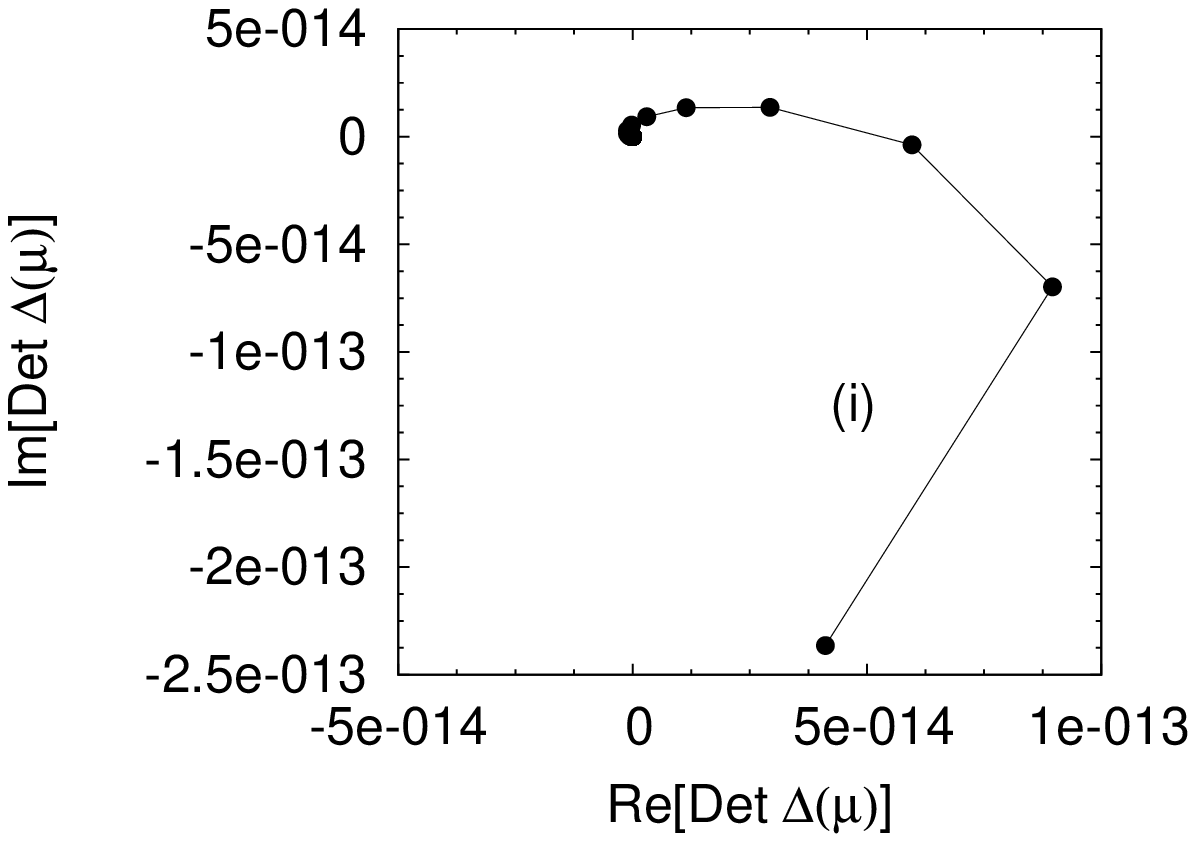}
\includegraphics[width=0.45\linewidth]{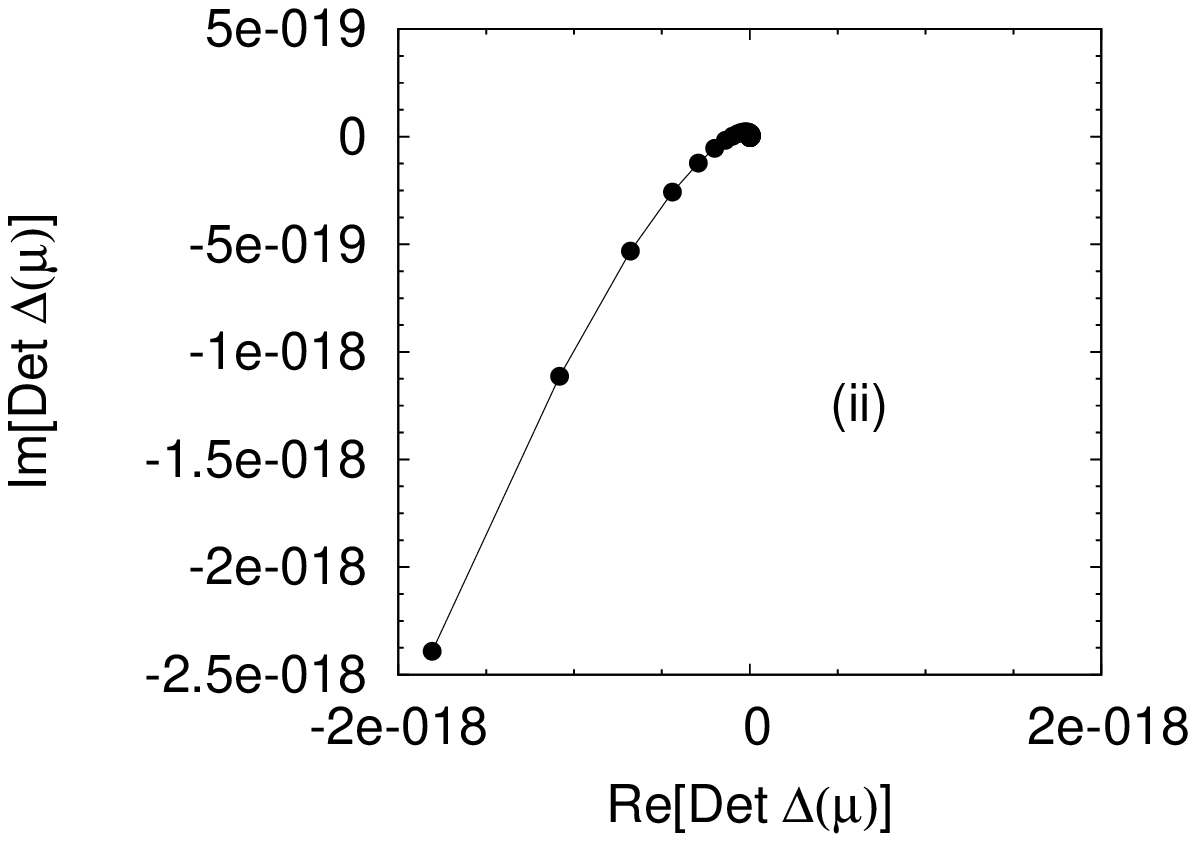}
\includegraphics[width=0.45\linewidth]{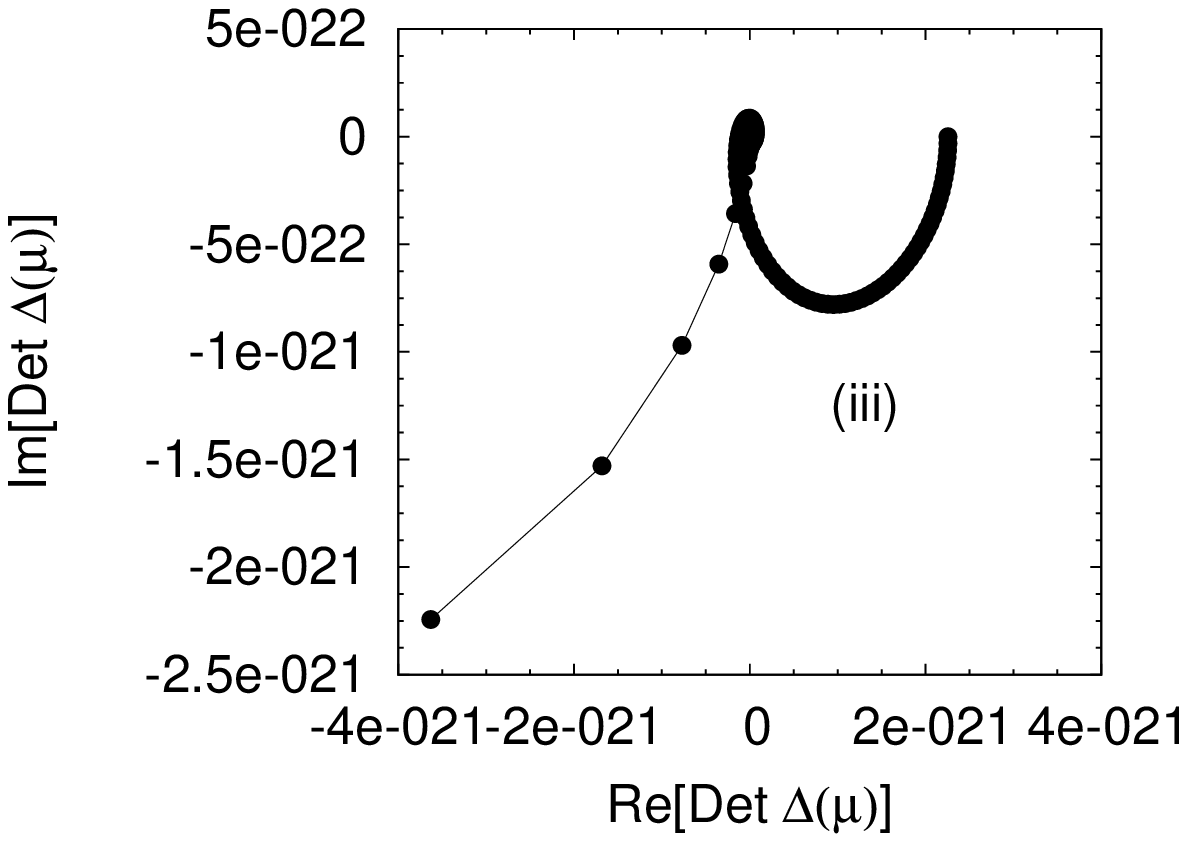}
\includegraphics[width=0.45\linewidth]{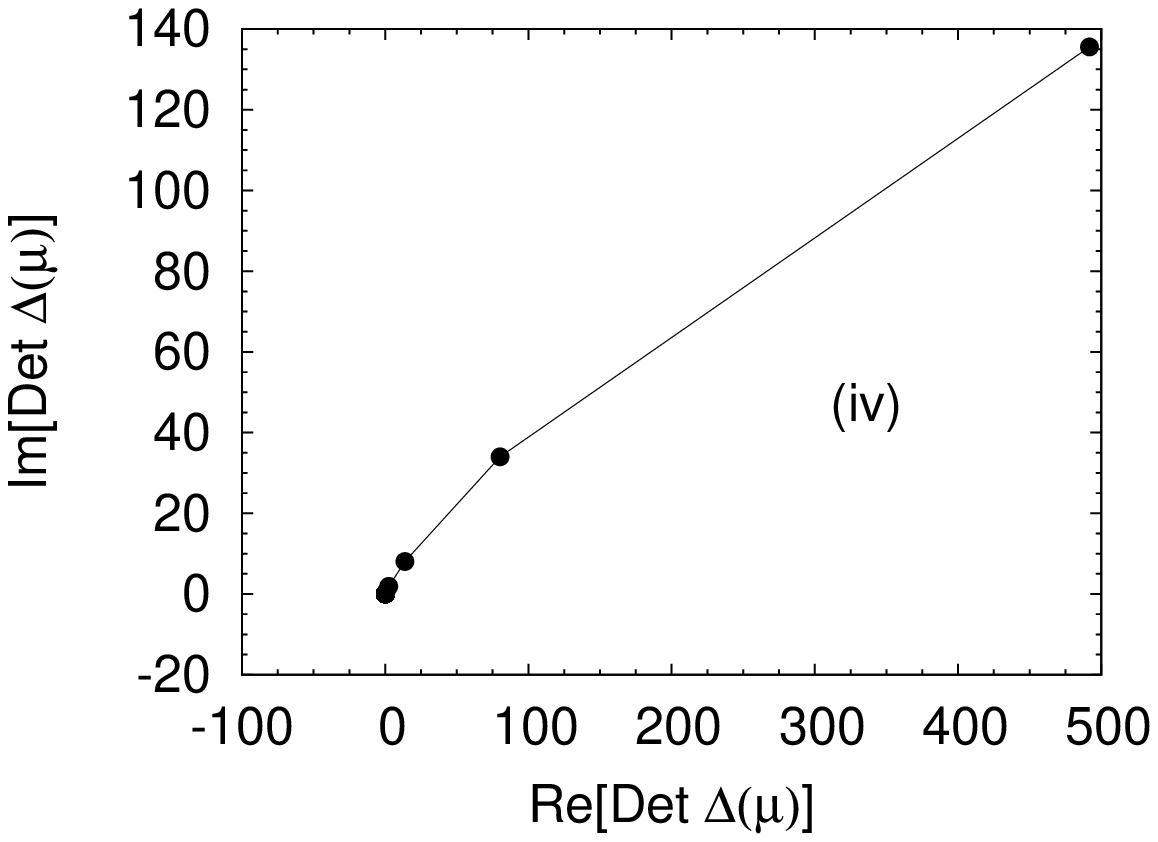}
\begin{minipage}{0.75\linewidth}
\caption{Parametric plot of $\Det \Delta(\mu)$ from $\mu=0$ to $\mu=1$
for $\beta = 1.85$. The points are denoted for $\delta \mu=0.01$.
}\label{Jul1810fig8}
\end{minipage}
\end{center}
\end{figure}
\begin{figure}[htbp]
\begin{center}
\includegraphics[width=0.45\linewidth]{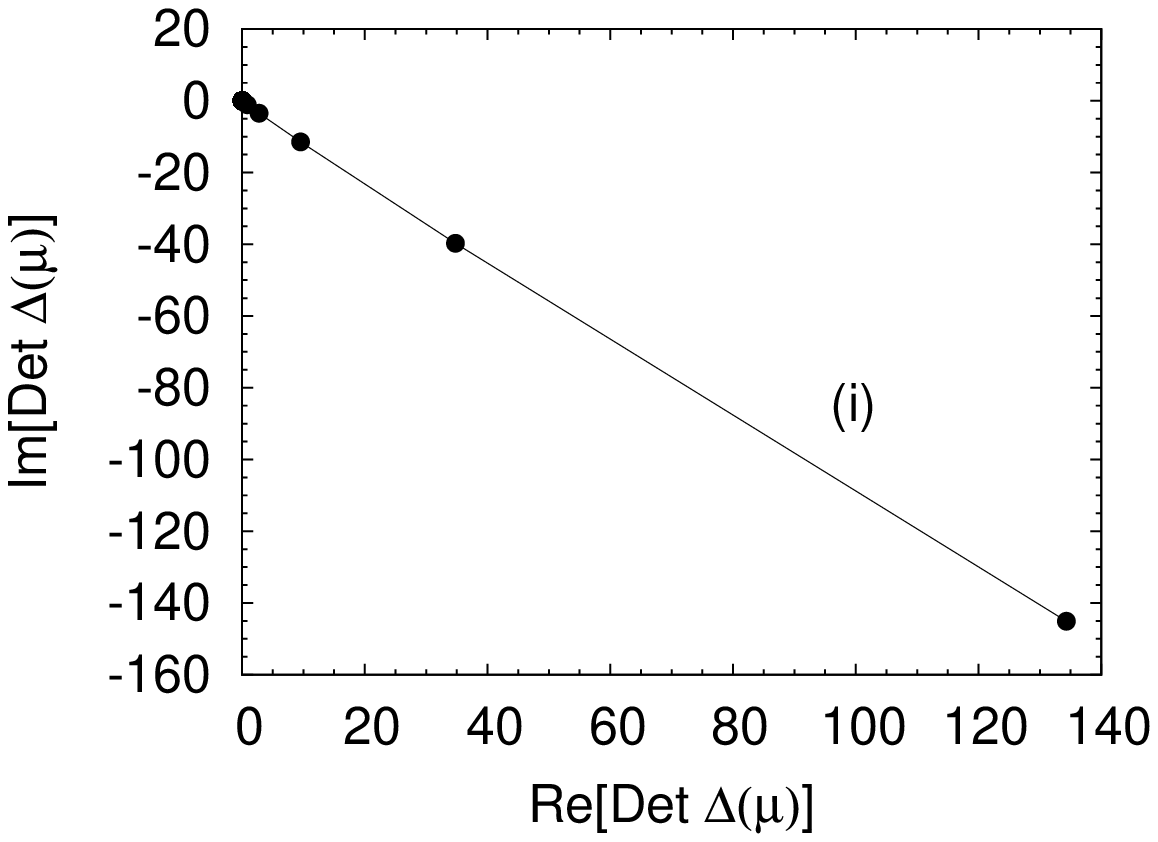}
\includegraphics[width=0.45\linewidth]{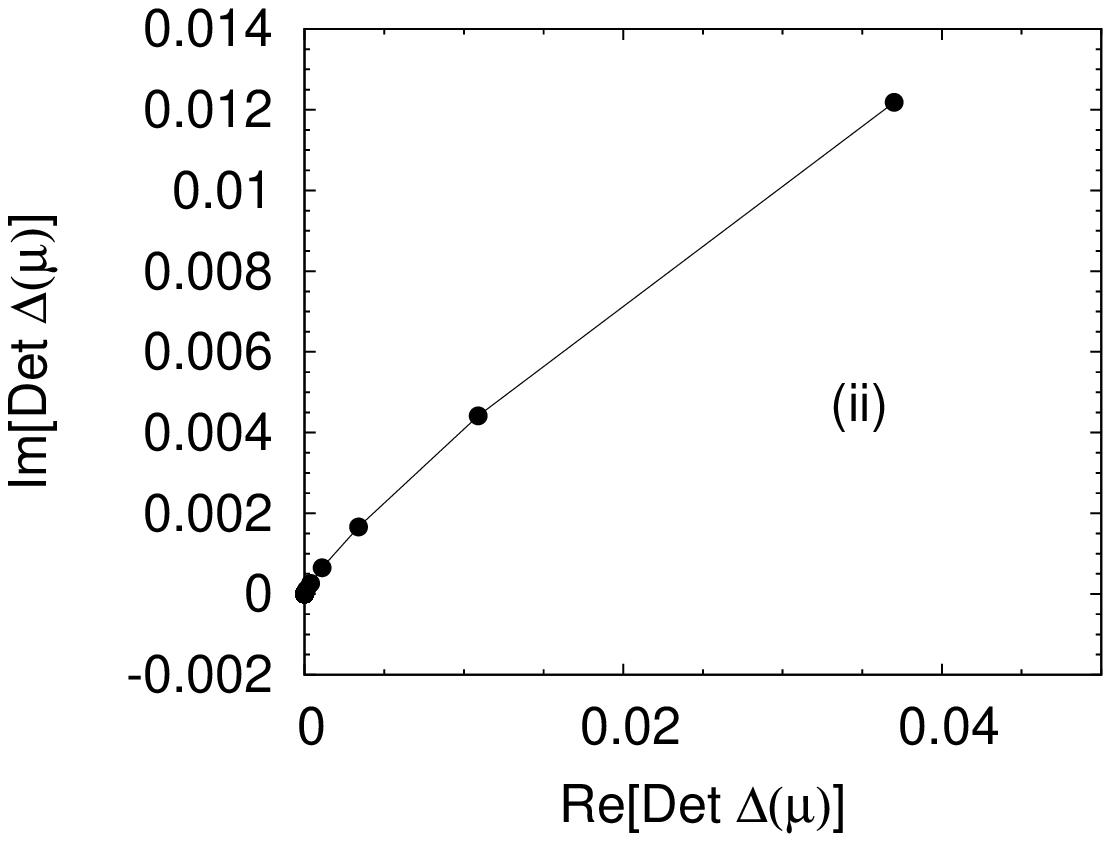}
\includegraphics[width=0.45\linewidth]{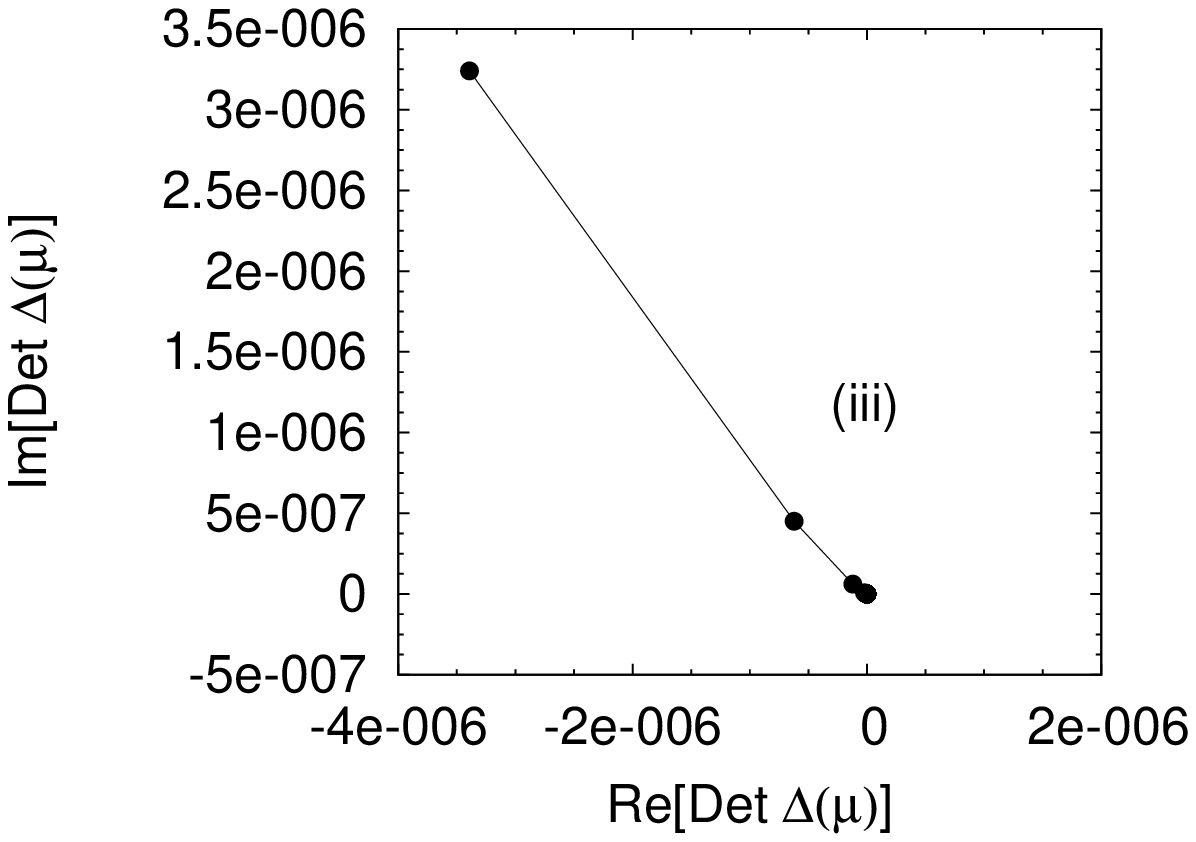}
\includegraphics[width=0.45\linewidth]{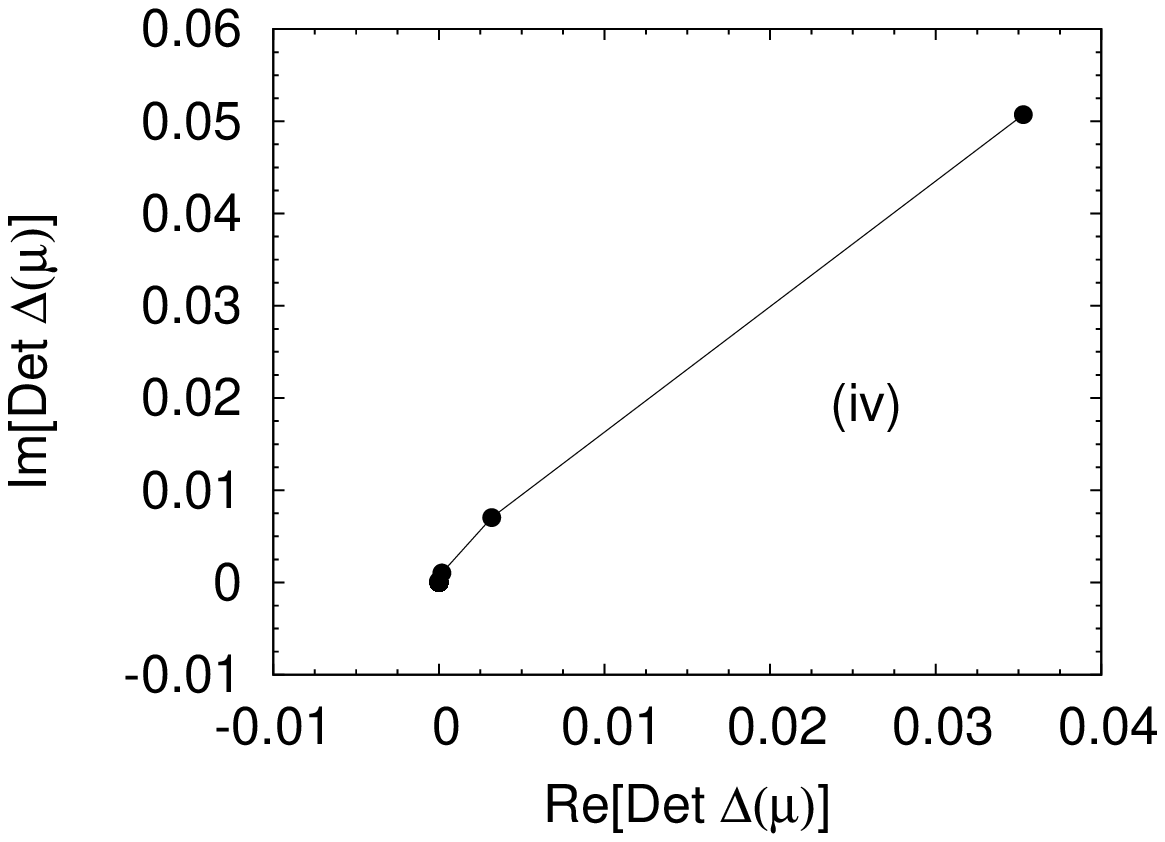}
\begin{minipage}{0.75\linewidth}
\caption{Parametric plot of $\Det \Delta(\mu)$ from $\mu=0$ to $\mu=1$
for $\beta = 2.0$. The points are denoted for $\delta \mu=0.01$.
}\label{Jul1810fig9}
\end{minipage}
\end{center}
\end{figure}
One of the advantages of the reduction method is that it makes easy to 
calculate the $\mu$ dependence of the quark determinant. Once we 
perform the reduction procedure and obtain $\lambda_n$ or $c_n$, 
we can obtain $\det \Delta(\mu)$ for arbitrary $\mu$. 
In Figs.~\ref{Jul1810fig8} and \ref{Jul1810fig9}, we show the $\mu$ 
dependence of the determinant. The values remain near the starting 
points when $\mu$ is small, and move rapidly when $\mu$ exceeds around 0.9.

\begin{figure}[htbp]
\begin{center}
\includegraphics[width=0.45\linewidth]{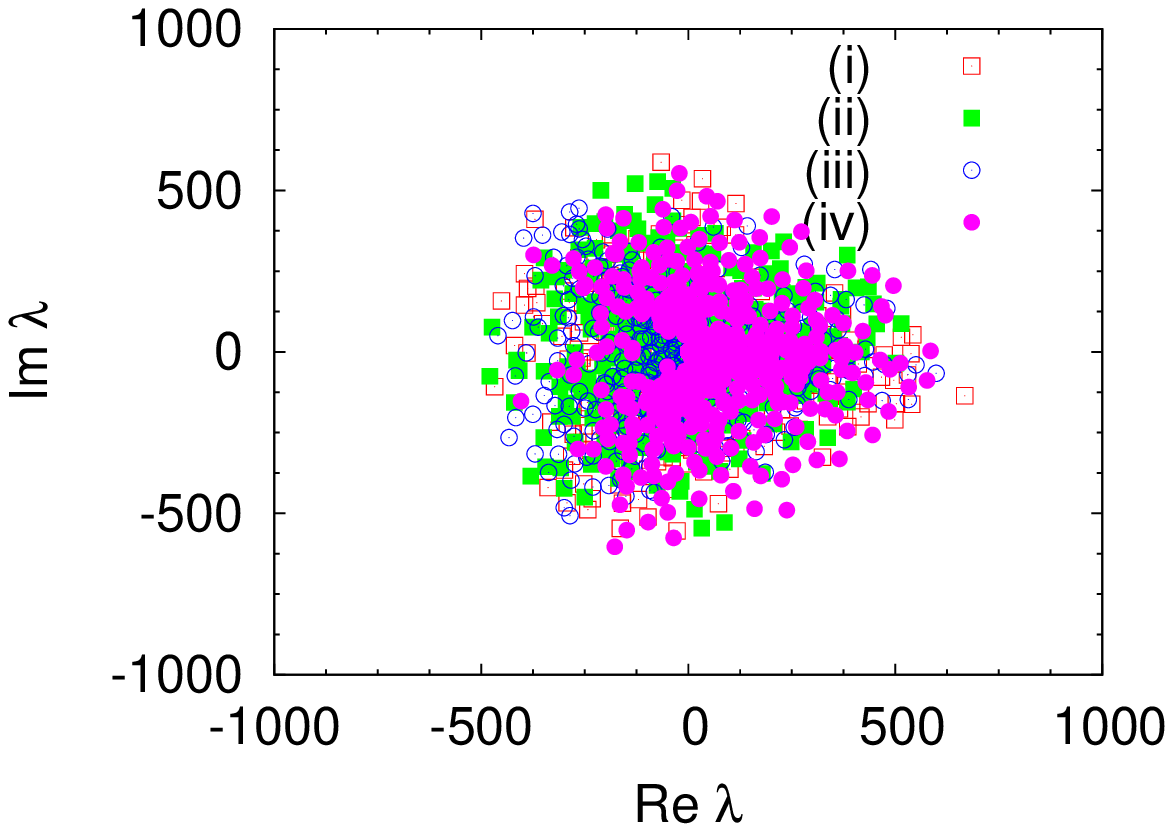}
\includegraphics[width=0.45\linewidth]{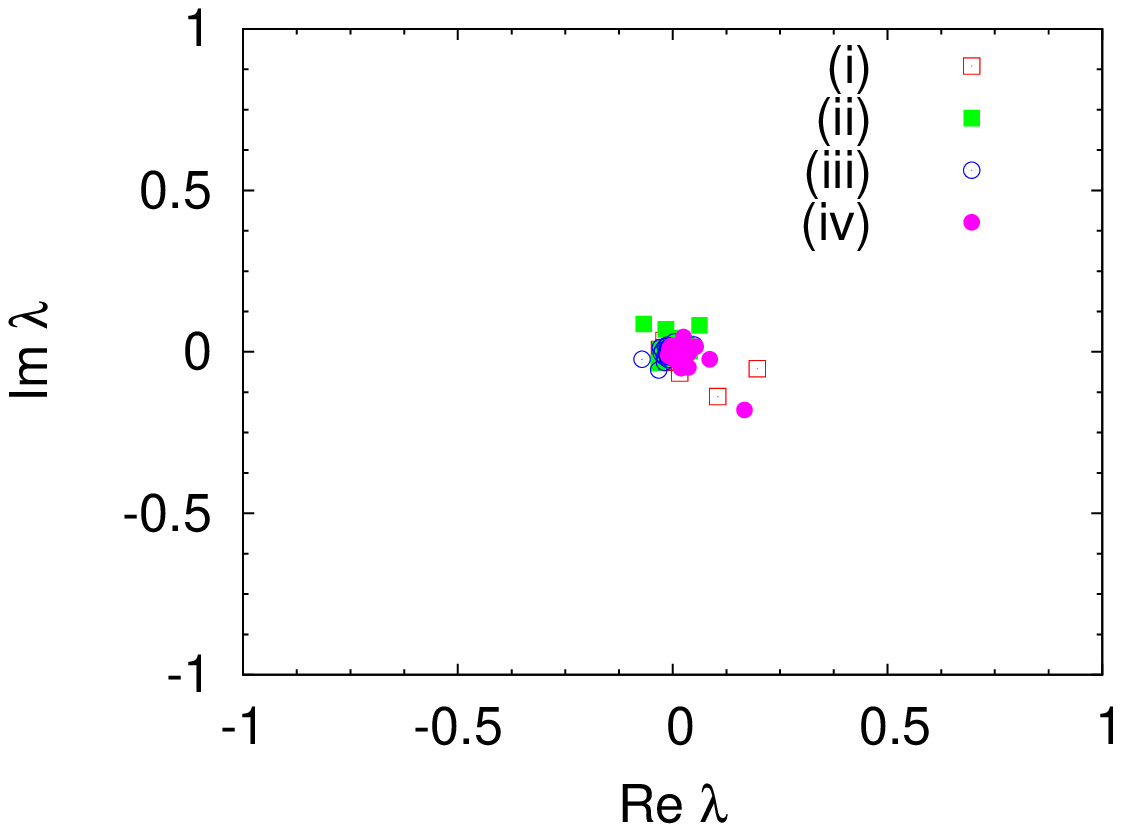}
\includegraphics[width=0.45\linewidth]{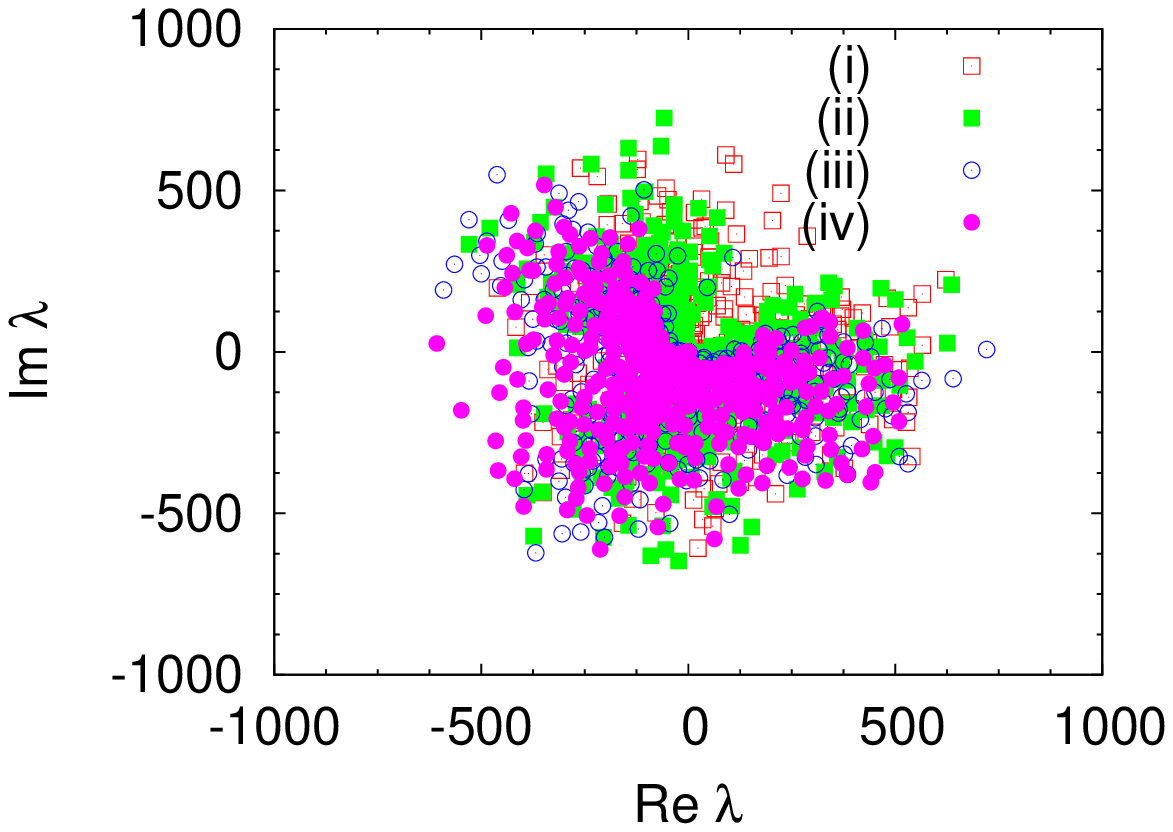}
\includegraphics[width=0.45\linewidth]{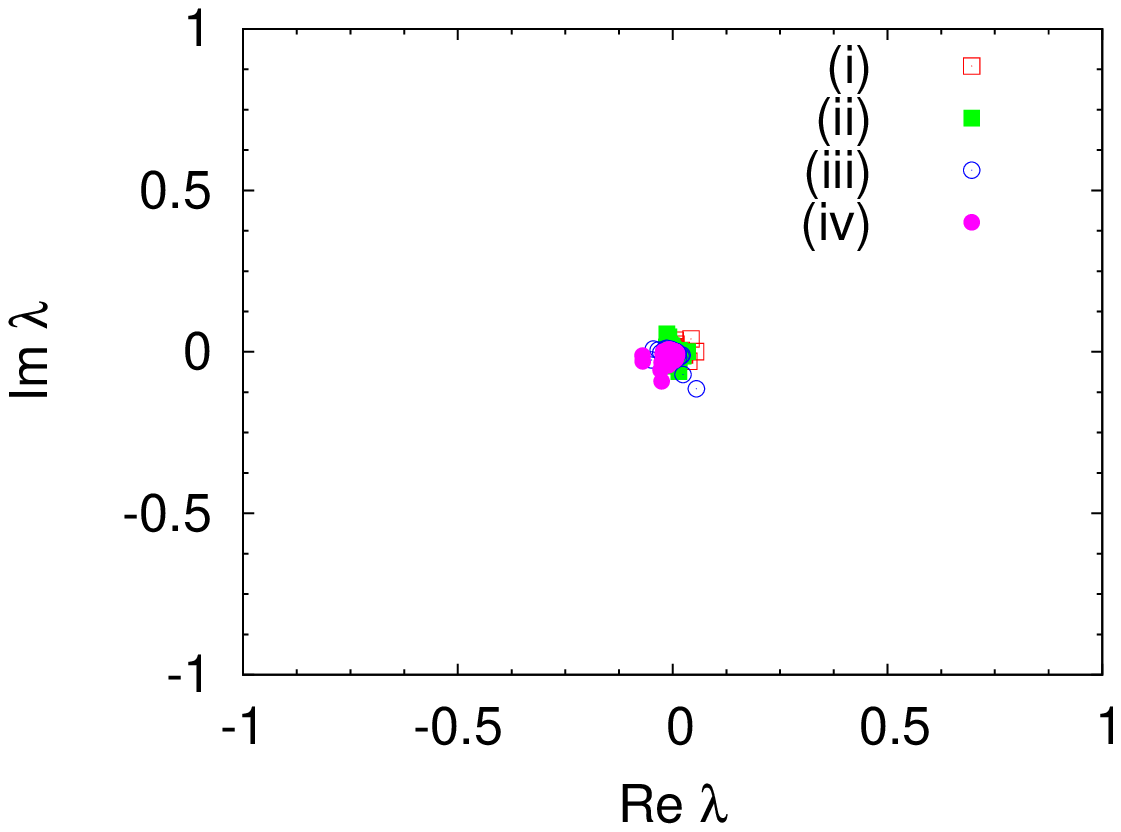}
\begin{minipage}{0.75\linewidth}
\caption{The distribution of the eigenvalues $\lambda$ in the complex 
plane. The topped and bottomed panels are for $\beta = 1.85$ and $2.0$, 
respectively. The left and right panels show the distributions in 
two different scales. 
}\label{Jul1810fig3}
\end{minipage}
\end{center}
\end{figure}
Next, we study the reduction method in more detail. 
The distribution of the eigenvalues $\lambda$ of the reduced matrix $Q$ 
is shown in Figs.~\ref{Jul1810fig3}. We observe that the eigenvalues 
are split in two regions. Almost half of the eigenvalues are distributed 
in a region $|\lambda | \agt 5$, and the other half in a region $|\lambda| \alt 0.5$.
There is a margin between two regions, where no eigenvalue is found, 
as we can see in the right panels in  Fig.~\ref{Jul1810fig3}.
The splitting of the eigenvalues is observed in the eight measurements. 
Note that the eigenvalues are constrained by the condition 
$\prod_{n=1}^{\Nred} \lambda_n =1$. Qualitatively, this can be 
understood from the fact that the matrix $Q$ is a product of the block 
matrices $A_i=(\alpha^{-1}_i \beta_i)$. It is expected that when the 
system is in equilibrium $A_i$ moderately depends on time. In such a 
case, we can express $A_i \sim \bar{A} + \delta A_i$, where $\bar{A}$ 
is independent of time. Assuming the time-dependent part $\delta A_i$ 
is small, $Q = \prod_{i=1}^{N_t} A_i \sim \bar{A}^{N_t} + {\cal O}(\delta)$. 
Then, $\bar{A}^{N_t}$ causes the splitting of the eigenvalues of the 
matrix $Q$ for ${\rm eigen}(\bar{A})>1$ and ${\rm eigen}(\bar{A})<1 $ cases. 

The coefficient $c_n$ is a polynomial of the eigenvalues $\lambda$
according to Eq.~(\ref{Eq:FugExpansion}). Because the number of the 
eigenvalues $\Nred$ is large, there appear two numerical problems. 
First problem is for an accuracy. We employ a recursive method in 
order to determine $c_n$ in enough precision. 
Second problem is that $c_n$ exceeds the range where a number can be represented
in double precision: about $10^{-308} \sim 10^{308}$. 
In order to overcome this problem, we develop a special routine to extend 
exponential part. See Appendix \ref{App:WideRangeNum}.
For the check, we compare our results with those obtained by using FM multi-precision
library (FMLIB)\cite{Web:FMLIB}. 

\begin{figure}[htbp]
\begin{center}
\includegraphics[width=0.45\linewidth]{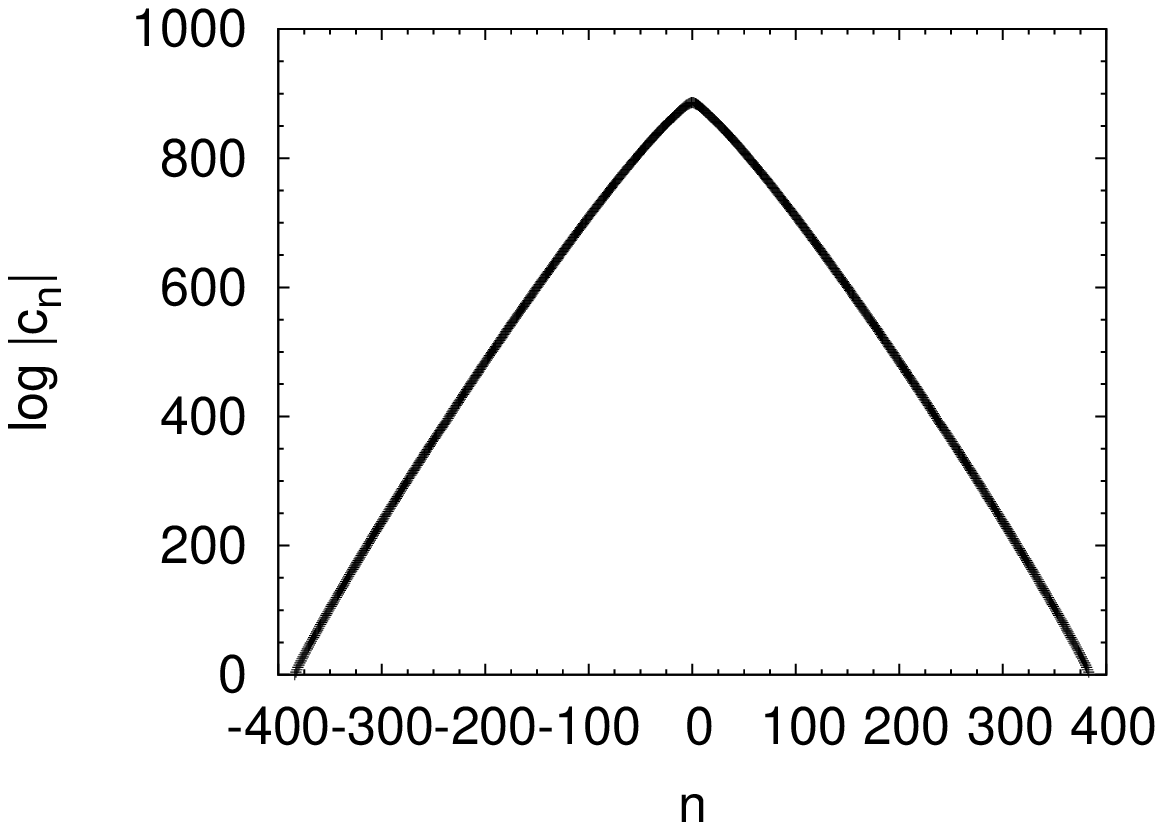}
\includegraphics[width=0.45\linewidth]{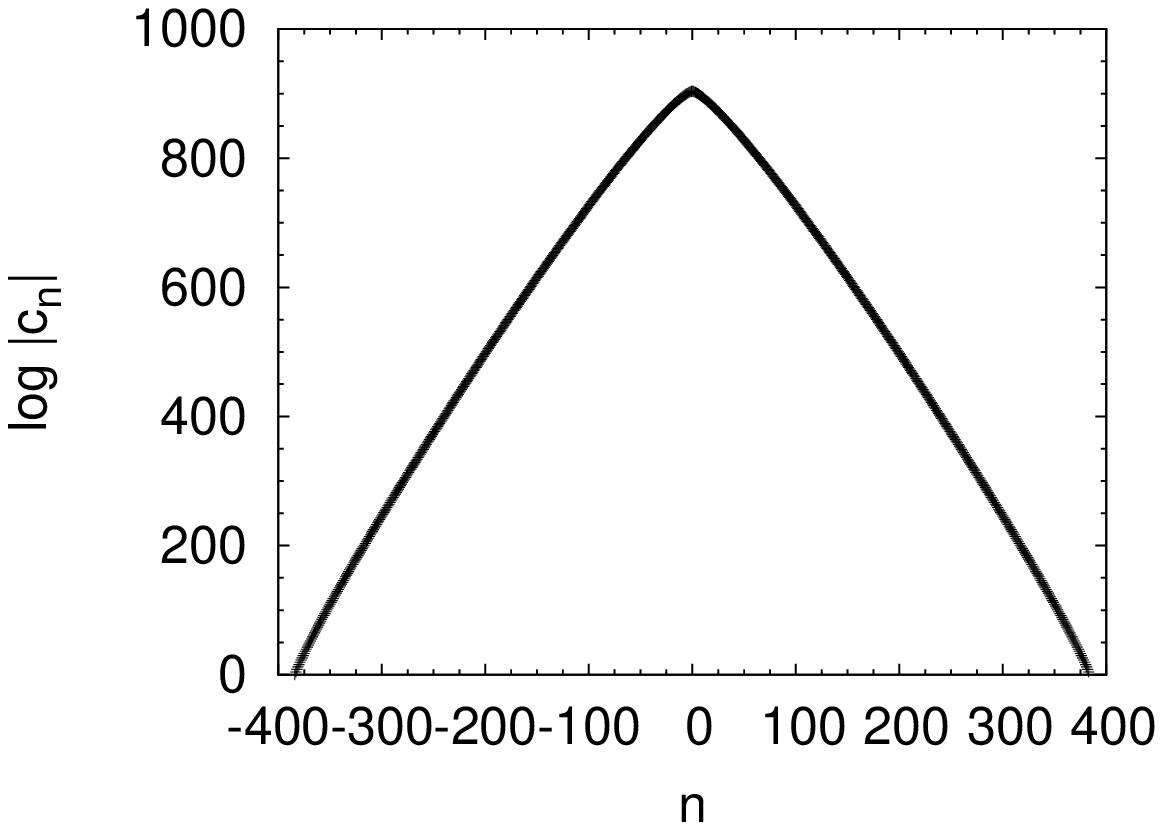}
\begin{minipage}{0.75\linewidth}
\caption{ The absolute values of $c_n$ in log scale. 
The left and right panels are for $\beta = 1.85$ and $2.0$, respectively. 
The results are obtained in the configurations (i). 
}\label{Jul1810fig5}
\end{minipage}
\end{center}
\end{figure}
We plot the absolute value of $c_n$ as a function of the winding number 
$n$ in Figs.~\ref{Jul1810fig5}, where we show the results only for 
the configurations (i) both in high and low temperatures 
because results for the other configurations are very similar to Figs.~\ref{Jul1810fig5}.
As we have mentioned, $|c_n|$ goes over the standard numerical range 
and reaches at $10^{900}$ at most, which is much larger than the 
maximum value in double precision. Note that the overall factor 
$C$ is order $10^{-900}$, then the cancellation between $C$ and 
$c_n$ makes their product $C_n = C c_n$ ordinary order.  
For both $\beta=1.8$ and $2.0$, we find that $|c_n|$ is maximum at $n=0$ and 
decreases exponentially as $|n|$ becomes larger. 

\begin{figure}[htbp]
\begin{center}
\includegraphics[width=0.45\linewidth]{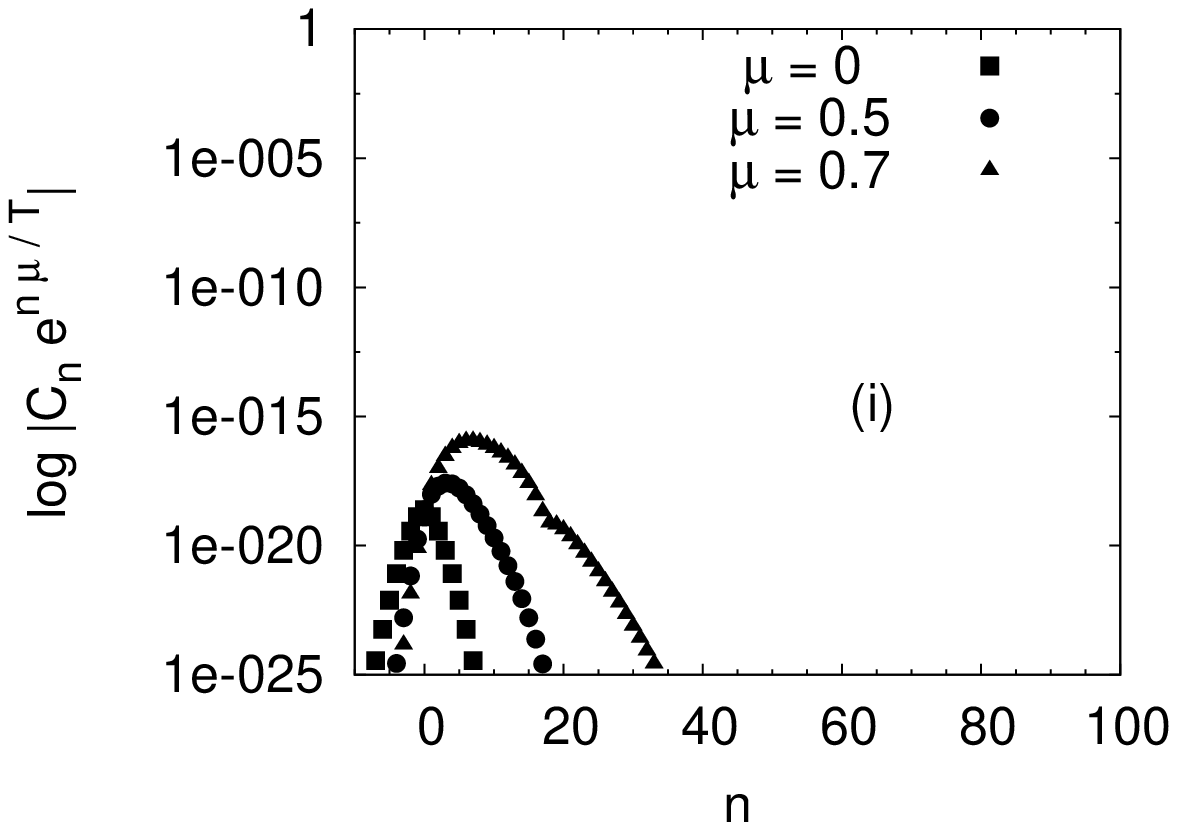}
\includegraphics[width=0.45\linewidth]{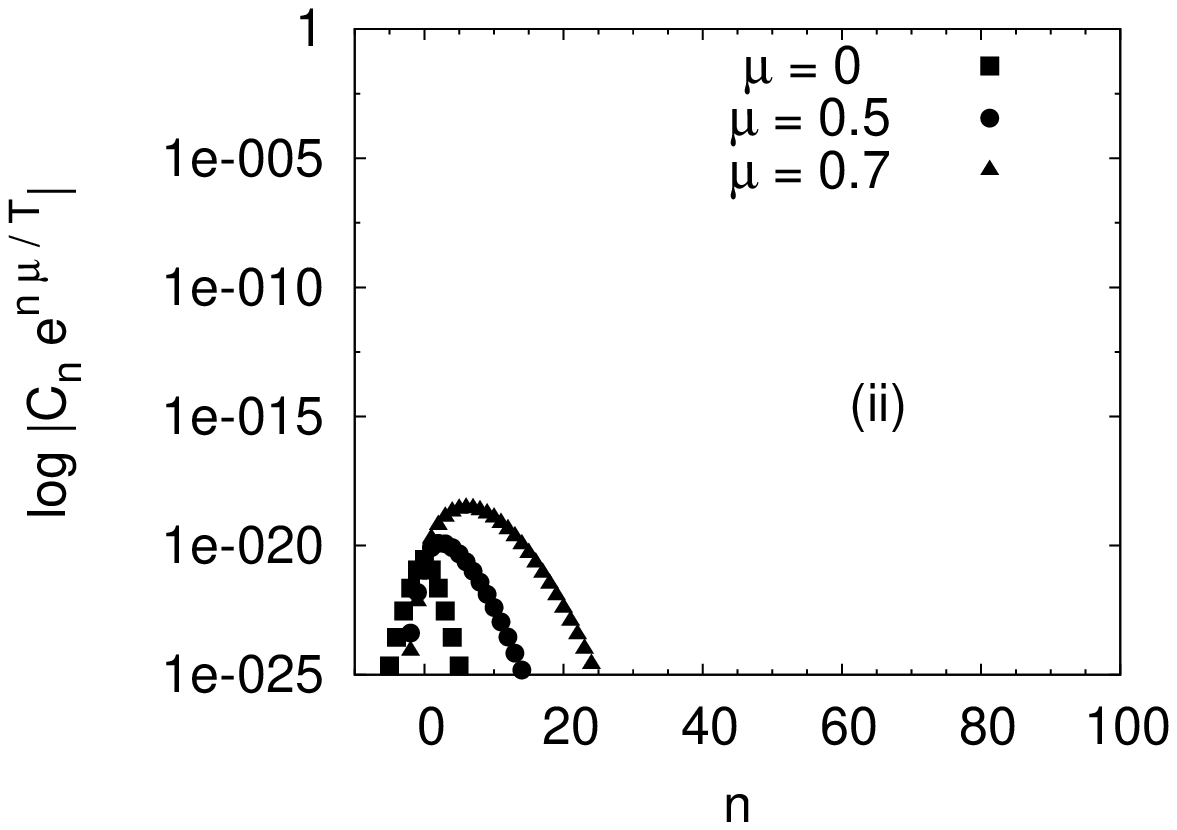}
\includegraphics[width=0.45\linewidth]{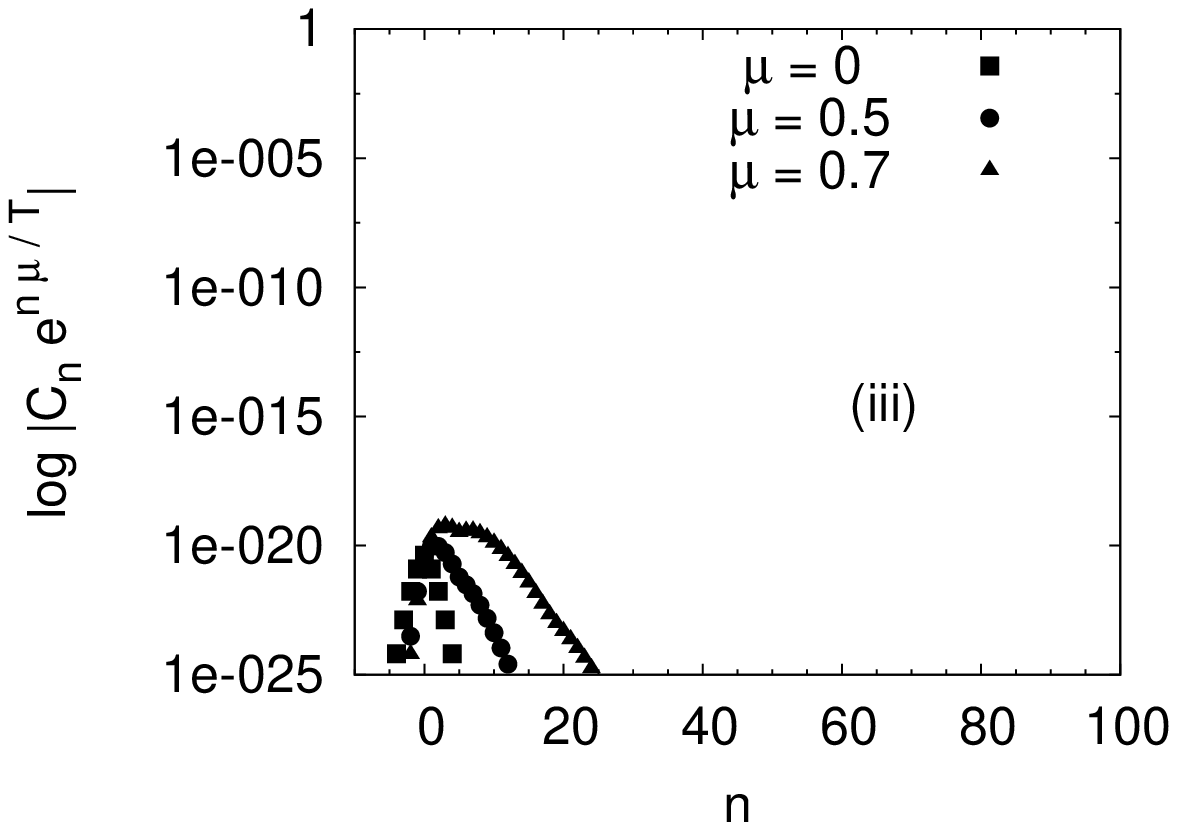}
\includegraphics[width=0.45\linewidth]{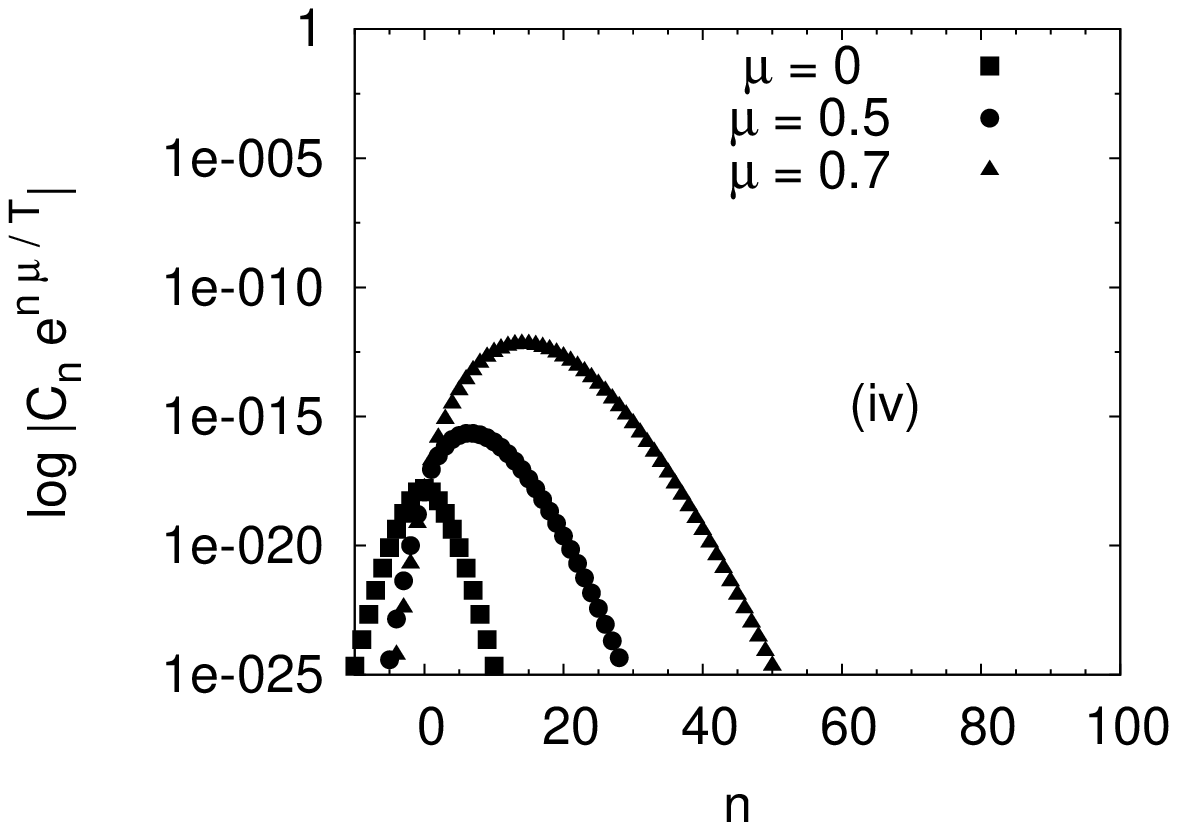}
\begin{minipage}{0.75\linewidth}
\caption{The distribution of $| C_n e^{n \mu /T}|$ for $\beta = 1.85$. 
The four panels correspond to the configurations (i), (ii), (iii) and (iv), respectively. 
}\label{Jul1810fig10}
\end{minipage}
\end{center}
\end{figure}

\begin{figure}[htbp]
\begin{center}
\includegraphics[width=0.45\linewidth]{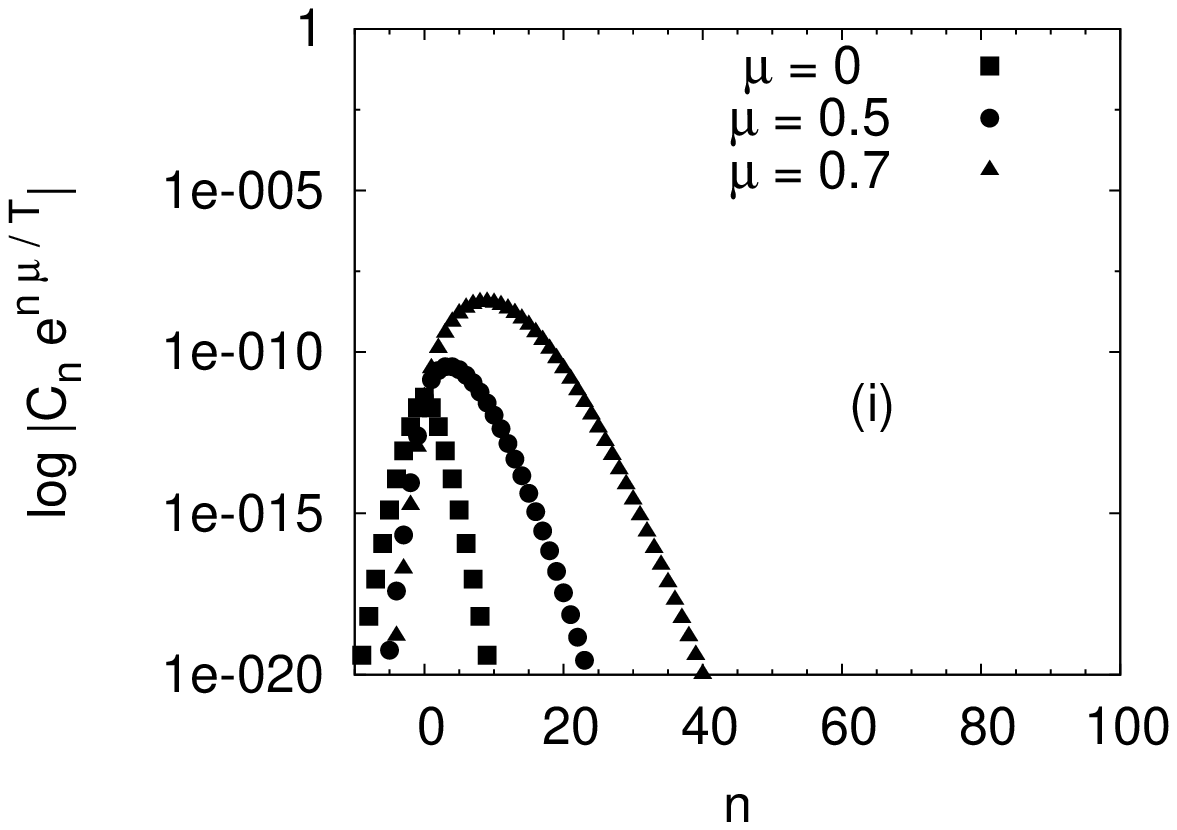}
\includegraphics[width=0.45\linewidth]{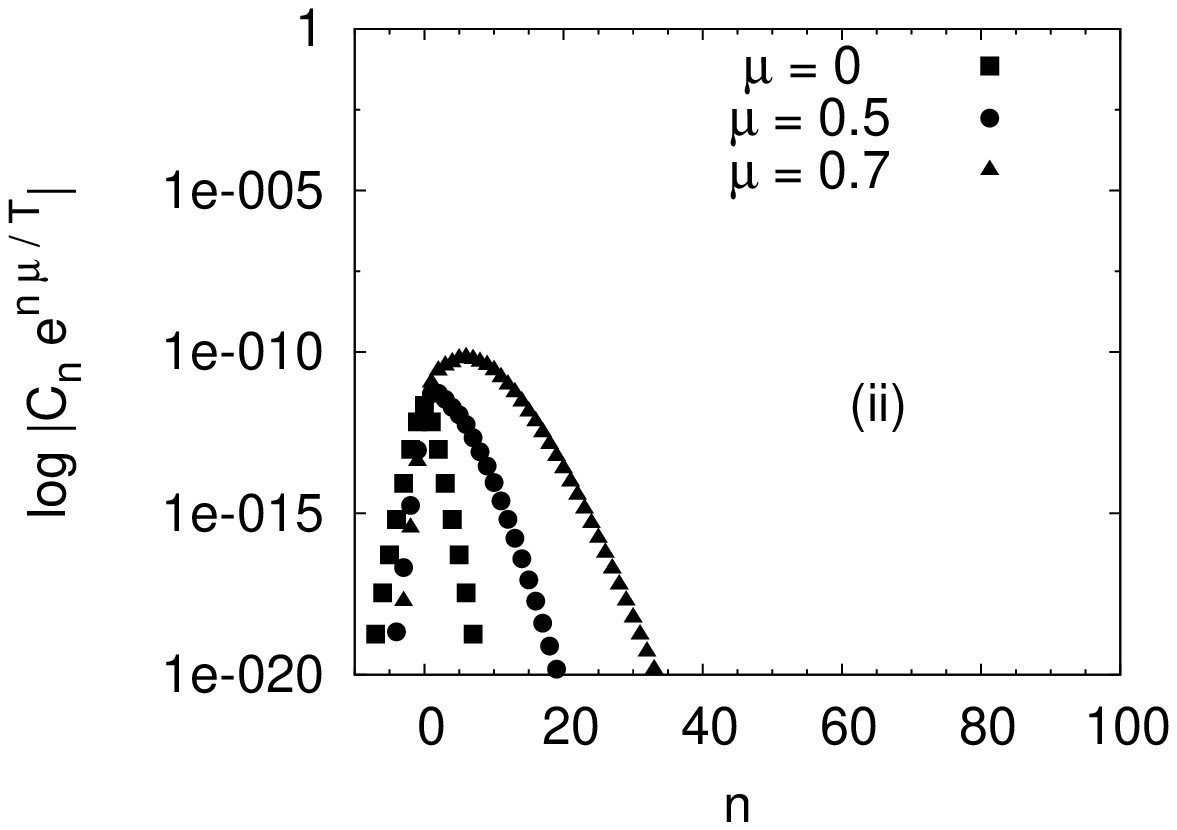}
\includegraphics[width=0.45\linewidth]{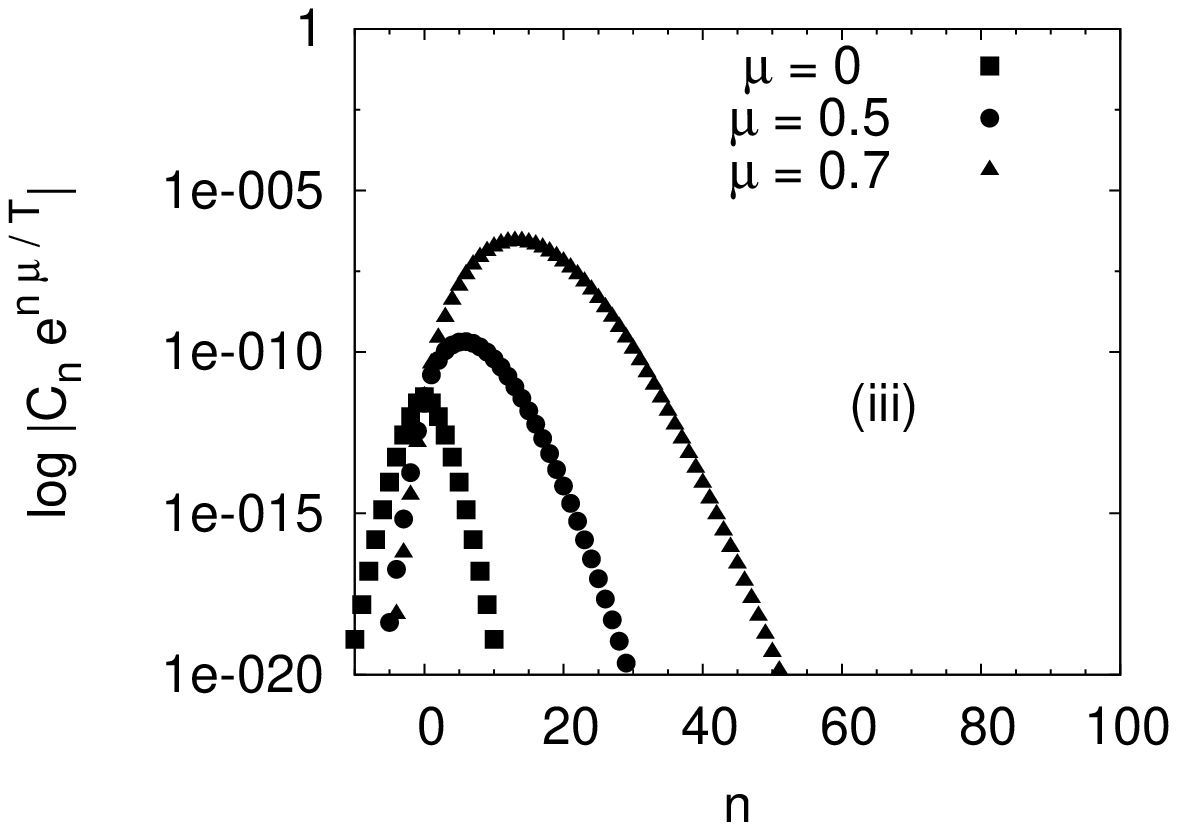}
\includegraphics[width=0.45\linewidth]{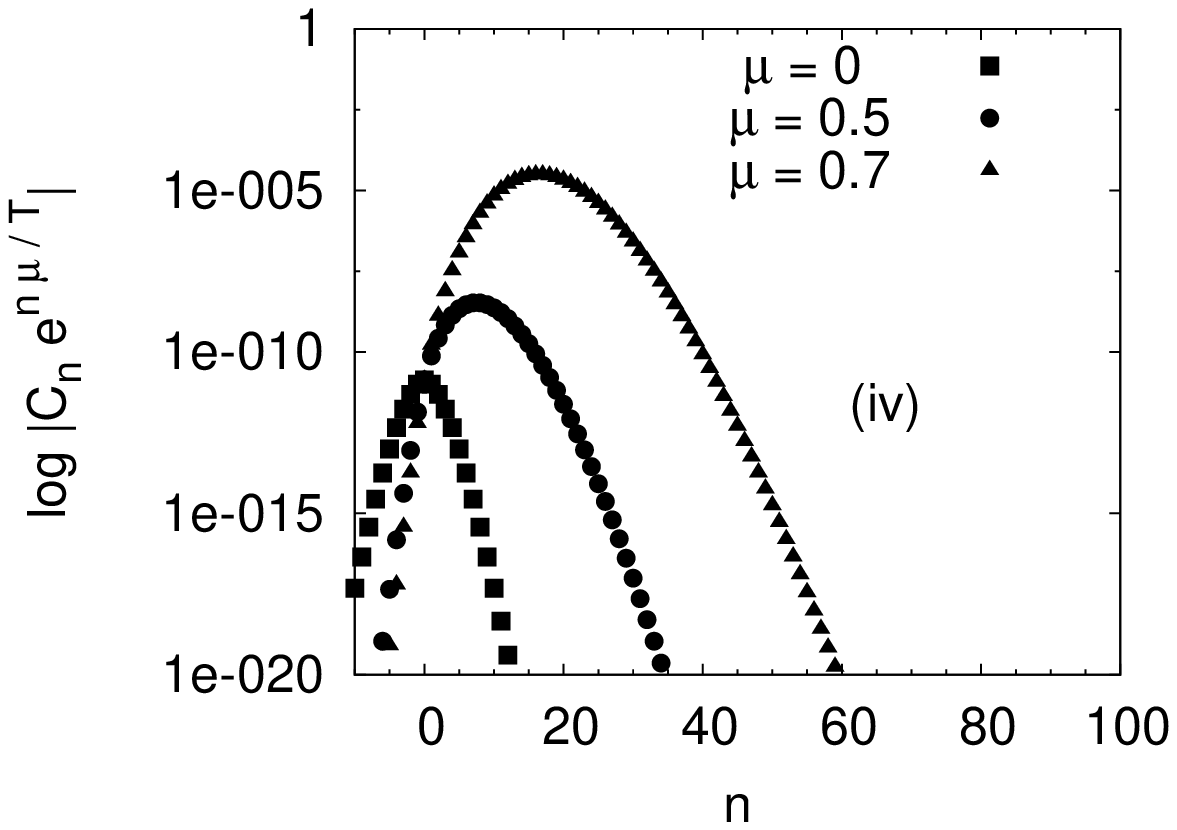}
\begin{minipage}{0.48\linewidth}
\caption{The distribution of $| C_n e^{n \mu /T}|$ for $\beta=2.0$. 
The four panels correspond to the configurations (i), (ii), (iii) 
and (iv), respectively. 
}\label{Jul1810fig11}
\end{minipage}
\end{center}
\end{figure}

Next, we show the absolute value of $C_n e^{n \mu/T}$ for several chemical 
potentials in Figs.~\ref{Jul1810fig10} and \ref{Jul1810fig11}. 
In contrast to $|c_n|$, the fugacity factor $e^{n \mu/T}$ becomes 
larger as $n$ becomes larger. 
The difference between the $n$-dependence of $|c_n|$ and $e^{n \mu/T}$ 
leads to a peak for $| C_n e^{n \mu/T} |$, as we can see in 
Figs.~\ref{Jul1810fig10} and \ref{Jul1810fig11}.
Several terms in the vicinity of the peak dominate $\det \Delta(\mu)$. 
For instance, $\det \Delta(0)$ is dominated by terms near $n=0$.
The location of the peak moves towards larger values of $n$ as $\mu$
 becomes larger. However the $\mu$ dependence of the location of 
the peak is not so strong. Even for the chemical potential near to $\mu =1$, 
significant contributions come from terms with $n<100$. 
In the following, we consider terms with $n<100$. 
 
\begin{figure}[htbp]
\begin{center}
\includegraphics[width=0.45\linewidth]{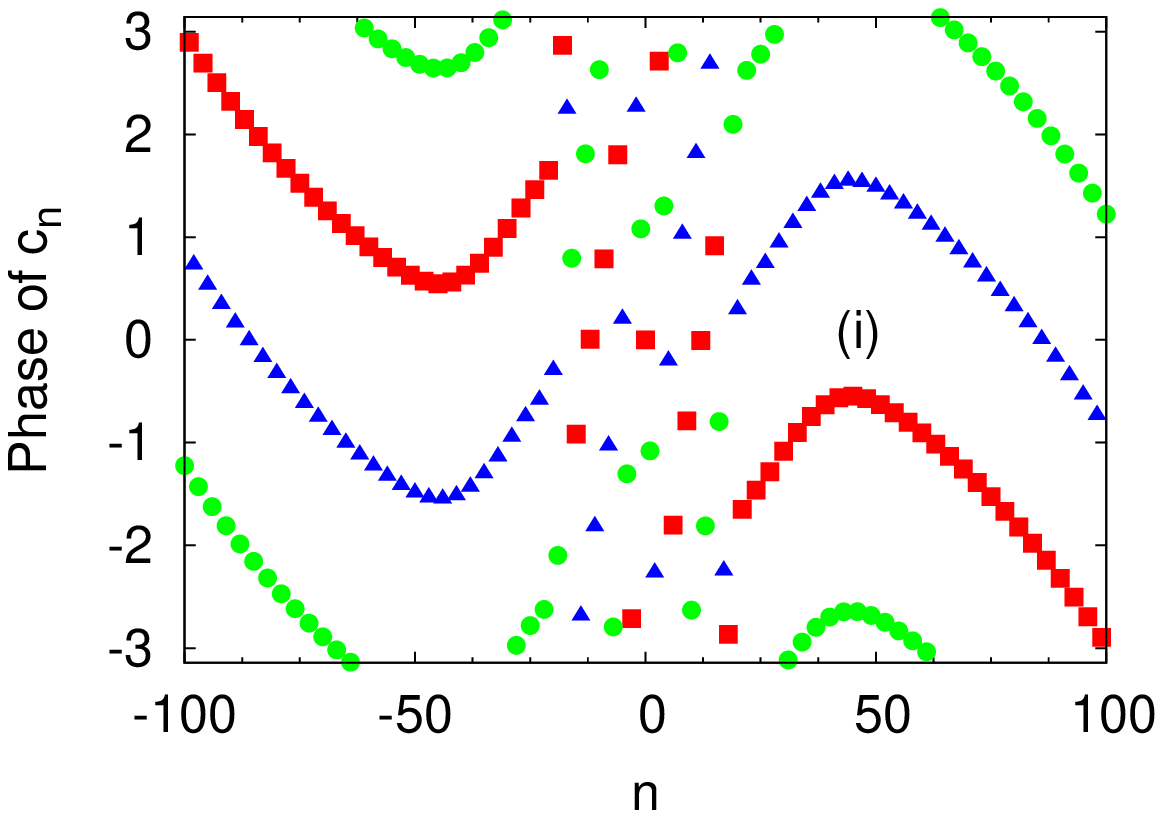}
\includegraphics[width=0.45\linewidth]{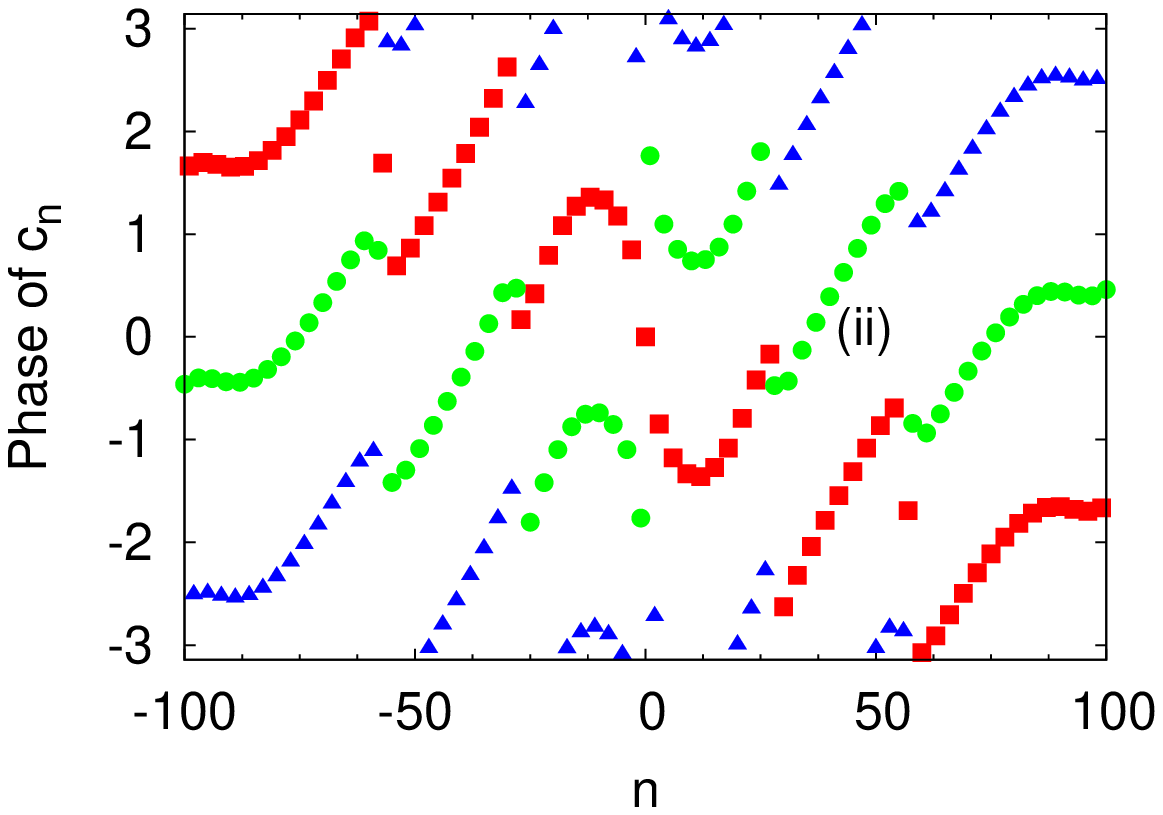}
\includegraphics[width=0.45\linewidth]{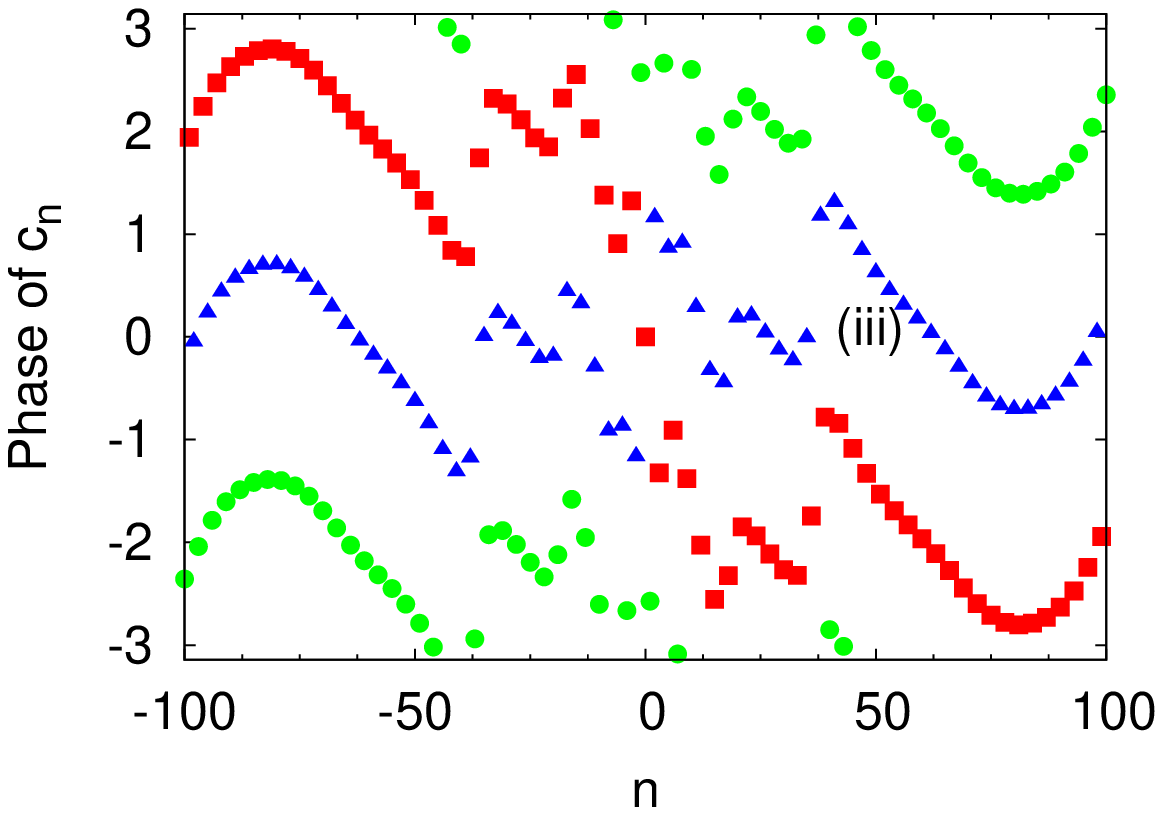}
\includegraphics[width=0.45\linewidth]{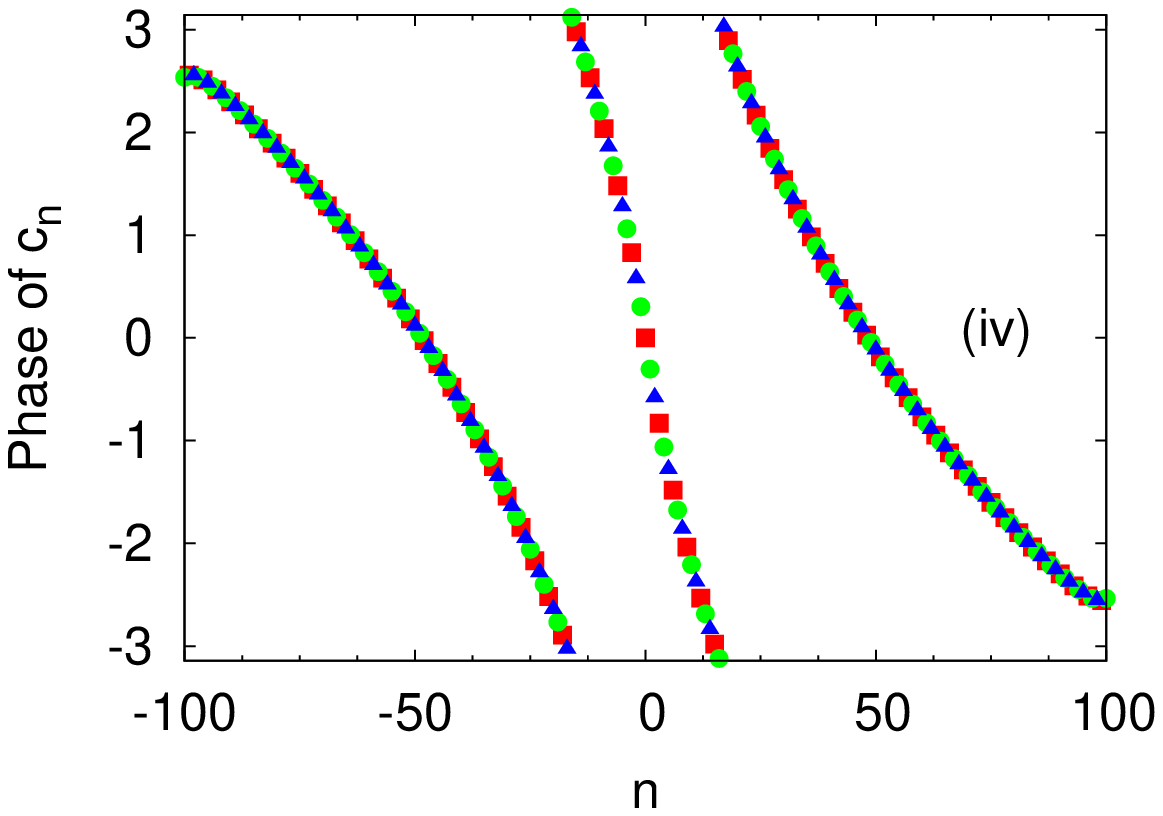}
\begin{minipage}{0.48\linewidth}
\caption{The phase of $c_n$ as functions of $n$ for $\beta = 1.85$. 
The four panels correspond to the configurations (i), (ii), (iii) and (iv), 
respectively. In each panel, square, circle and triangle denote $c_n$ for 
mod$(n,3)$=0, 1 and 2, respectively. 
}\label{Jul1810fig6}
\end{minipage}
\end{center}
\end{figure}

\begin{figure}[htbp]
\begin{center}
\includegraphics[width=0.45\linewidth]{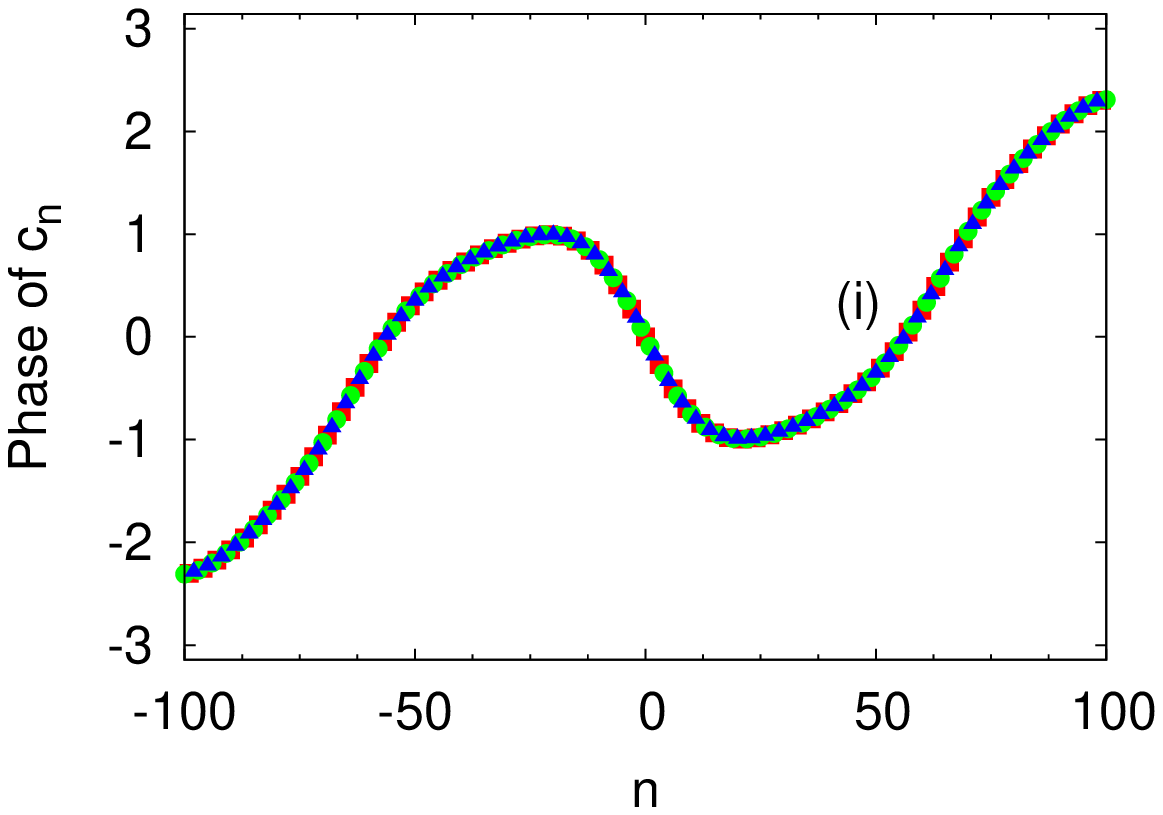}
\includegraphics[width=0.45\linewidth]{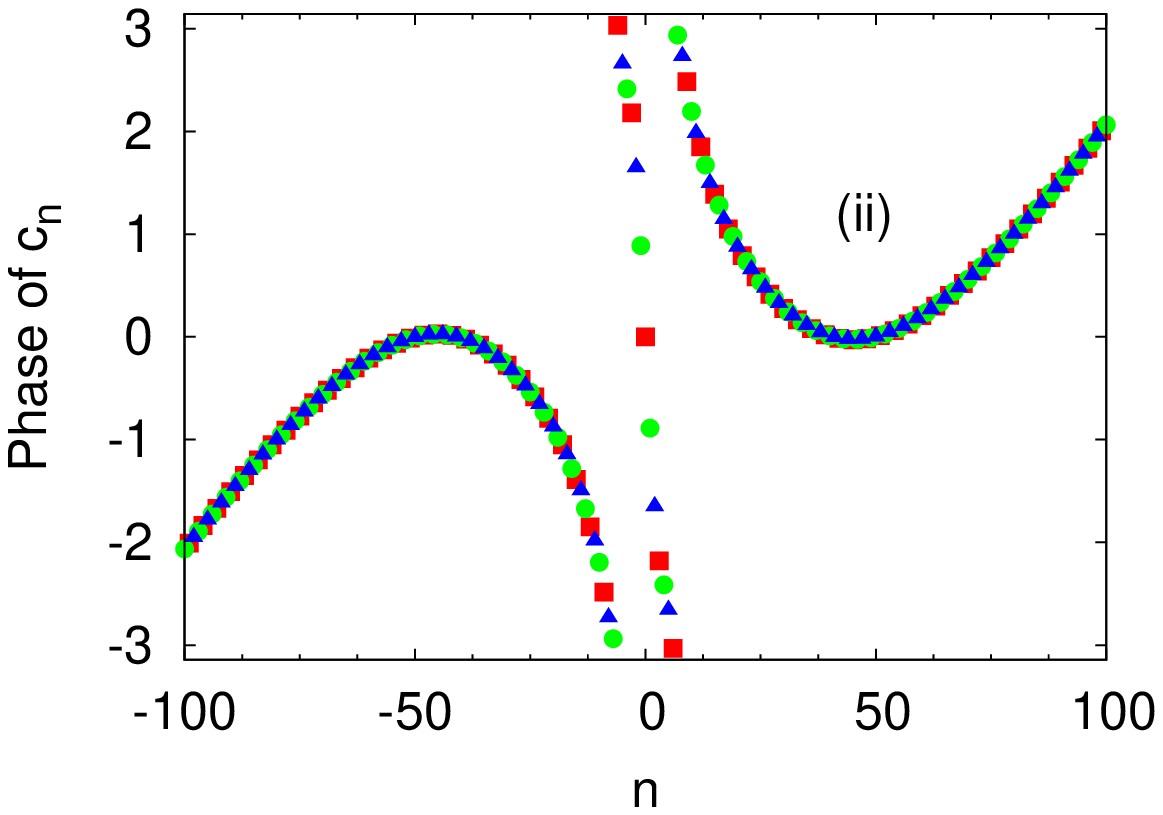}
\includegraphics[width=0.45\linewidth]{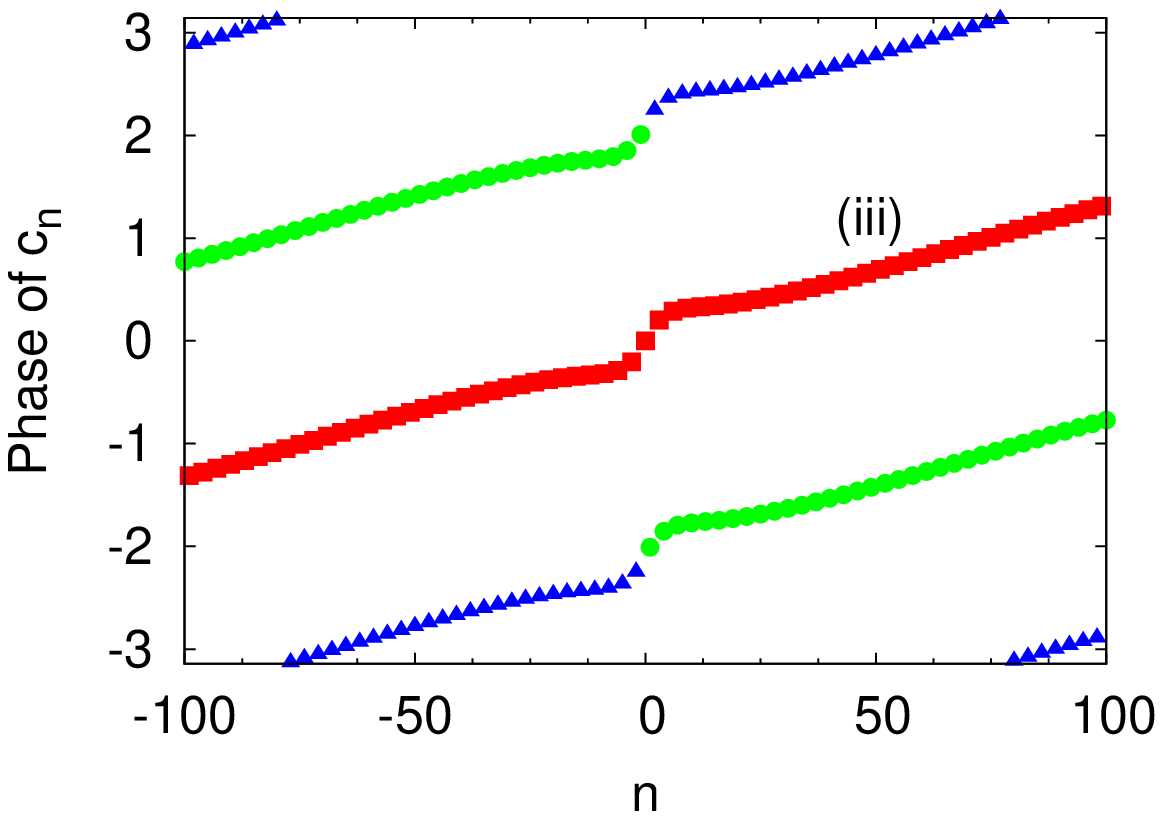}
\includegraphics[width=0.45\linewidth]{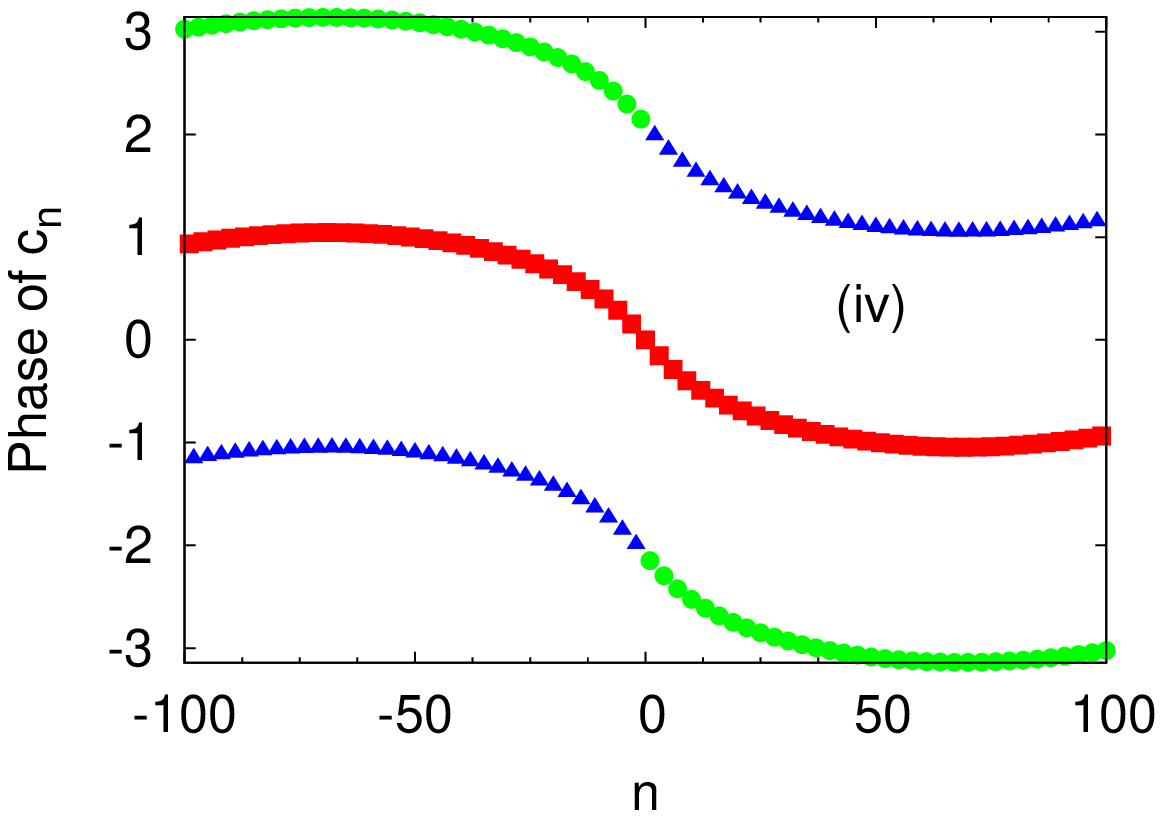}
\begin{minipage}{0.48\linewidth}
\caption{The phase of $c_n$ as functions of $n$ for $\beta = 2.0$. 
The four panels correspond to the configurations (i), (ii), (iii) and (iv), 
respectively. In each panel, square, circle and triangle denote $c_n$ for 
mod$(n,3)$=0, 1 and 2, respectively. 
}\label{Jul1810fig7}
\end{minipage}
\end{center}
\end{figure}
The phase of $c_n$ as a function of $n$ are shown in Figs.~\ref{Jul1810fig6} 
and Figs.~\ref{Jul1810fig7}. 
In all the eight configurations, there is a symmetry under a rotation 
with $\pi$. This symmetry indicates the relation $c_n^* = c_{-n}$ is numerically 
satisfied. 
The phase of $c_n$ complicatedly depends on the winding number, temperature 
and configuration. We find that there are two particular $n$-dependence. 
One is that the phase of $c_n$ is a continuous function of $n$, e.g. 
see (iv) in Figs.~\ref{Jul1810fig6}. 
(This means that we can fit the phase of $c_n$ in terms of a 
continuous function, although $n$ is not a continuous variable.)
The other is that the phase of $c_n$ is spit into three lines 
classified in terms of mod$(n,3)$, e.g. see (i) in Figs.~\ref{Jul1810fig6}. 
Each line is a continuous function of $n=3m, 3m+1$ or $3m+2$ 
and there is a gap about $2\pi /3$ between lines. 
This splitting causes the cancellation of $C_n e^{n \mu/T}$ among 
neighboring three terms with $n=3m, 3m+1, 3m+2$ and suppresses 
the magnitude of the determinant. For instance, the four values of 
$\det \Delta(\mu)$ corresponding to the four configurations in $\beta=1.85$ 
are classified in two groups, as we have seen in Figs.~\ref{Jul1810fig8}; 
$\det \Delta(\mu)$ is of order $10^{-20}$ in the configurations (i), 
(ii) and (iii), and it is of order $100$ in the configuration (iv). 
This observation is related to the behavior of the phase of $c_n$. 

\section{Summary}
In this paper, we have presented the reduction formula for the Wilson fermion 
determinant. The formula reduces the numerical cost to evaluate the Wilson fermion 
determinant. 
The point is that the Wilson fermion matrix contains the projection operators, 
which enable to transform the fermion matrix so that the temporal part 
of the determinant can be performed analytically. Thus the Wilson fermion 
determinant is reduced to the determinant of the reduced matrix. 
Solving the eigenvalue problem for the reduced matrix, 
the determinant is expressed in powers of fugacity. 
Although the basic idea for the reduction method is similar to that for 
staggered fermions, a difference comes from the use of the projection 
operators.

We perform the numerical simulations on the $4^4$ lattice and calculate 
the Wilson fermion determinant using the reduction formula. 
In order to determine the coefficients of the fugacity expansion 
in enough accuracy, we employ the recursive method and develop the 
special routine. Furthermore, we compare our results with those 
obtained by using a multi-precision library. 

We discussed the properties of the eigenvalues of the reduced matrix and 
of the coefficients $c_n$ of the fugacity expansion. 
The eigenvalues show an interesting behaviour; they are split in two regions. 
We find that there are two particular behaviours for the winding number dependence 
of the phase of the coefficients. One is that the phase of $c_n$ is
a continuous function of the winding number. The other is that the phase of $c_n$ is split into 
three lines classified in terms of mod$(n,3)$ with the gap about $2\pi /3$
between lines. 

\section*{Acknowledgment}

We thank Philippe de Forcrand and Andrei Alexandru for very useful
discussions, without which this work could not be completed.

After finishing this work, we noticed that Urs Wenger and Andrei Alexandru 
have developed a similar formula for the Wilson fermions. We thank them for
the correspondence.

The simulation was performed on NEC SX-8R at RCNP, and NEC SX-9 at CMC, 
Osaka University. We appreciate the warm hospitality and support of the 
RCNP administrators. This work was supported by Grants-in-Aid for 
Scientific Research 20340055 and 20105003.
\clearpage
\appendix
\section{Determinant of permutation matrix}
\label{Apr1010sec1}
Here we calculate the determinant of the permutation matrix $P$. 
Since the projection operators are singular $\det (r_\pm) =0$, 
we need to use a simple trick in order to obtain $\det P$; first we
reduce the determinant in the case that $r\neq 1$.  
Then we take the limit $r \to 1$, after eliminating the singularity.
To perform this, we summarize identities of the projection operators for 
arbitrary $r$. They are defined by  
\begin{align}
r_{\pm} = \frac{r \pm \gamma_4}{2}.
\end{align}
Using the definitions, it is straightforward to obtain  
\begin{subequations}
\begin{align}
r_+ r_- = r_- r_+ = \frac{r^2 -1 }{4} = \epsilon.
\end{align}
These equations lead to the inverse matrices 
\begin{align}
(r_+)^{-1} = \frac{1}{\epsilon} r_-\;, \;\;
(r_-)^{-1} = \frac{1}{\epsilon} r_+ .
\end{align}
\label{Jun1810eq2}
\end{subequations}
Using Eqs.~(\ref{Jun1810eq2}), we obtain 
\begin{align}
\det P & = \left( \begin{array}{ccccc} 
c_a r_-  & c_b r_+ z^{-1} V_{1,2} &         &        &  \\
       & c_a r_-  & c_b r_+ z^{-1} V_{2,3}  &        &  \\
       &        & c_a r_- & \ddots &  \\
       &        &       & \ddots & c_b r_+ z^{-1} V_{N_t-1, N_t}\\
- c_b r_+ z^{-1} V_{N_t,1} &    &         &        &  c_a r_-
\end{array}\right) \nn \\
& = \det \left( c_a^{N_t} r_-^{N_t} + c_b^{N_t} r_+^{N_t} z^{-N_t} \prod_{t_i=1}^{N_t} V_{t_i,t_i+1}  \right)
\label{Jun1810eq3}
\end{align}
Considering the Dirac components, $(r_\pm)^{N_t}$ are given by 
\begin{subequations}
\begin{align}
(r_+)^{N_t} &= \left( \begin{array}{cccc}
\left(\frac{r+1}{2}\right)^2 & & & \\
& \left(\frac{r+1}{2}\right)^2 & & \\
& & \left(\frac{r-1}{2}\right)^2 & \\
& & & \left(\frac{r-1}{2}\right)^2
\end{array}\right),  \\
(r_-)^{N_t} &= \left( \begin{array}{cccc}
\left(\frac{r-1}{2}\right)^2 & & & \\
& \left(\frac{r-1}{2}\right)^2 & & \\
& & \left(\frac{r+1}{2}\right)^2 & \\
& & & \left(\frac{r+1}{2}\right)^2
\end{array}\right).
\end{align}
\end{subequations}
Having these two terms, there is no singularity in Eq.~(\ref{Jun1810eq3}).
Then, we obtain 
\begin{align}
\det P = z^{-N/2} (c_a c_b)^{N/2}.
\end{align}

\section{Calculation of coefficients $c_n$}\label{App:WideRangeNum}

As we have shown, the coefficients $c_n$ in Eq.~(\ref{Eq:FugExpansion}) vary 
from order one to order $10^{900}$ even on the small $4^4$ lattice. They 
cannot be handled in the double precision. 
This problem usually happens when we consider expansions of the 
fermion determinant. 
So far, arbitrary accuracy libraries are often employed in order to 
calculate the coefficients.

We calculate $c_n$ as follows in a recursive way:
\be
\sum_{k=0}^M C_k'\xi^k
= 
(B_0 + B_1 \xi ) \sum_{k=0}^{M-1} C_k\xi^k
\ee
and
\bea
C_0' &=& B_0 C_0
\\
C_k' &=& B_{k-1} C_k + B_k C_{k-1}
\quad  (k = 1, 2, \cdots, M-1)
\label{Eq:recursive}
\\
C_M' &=& B_1 C_{M-1}
\eea
In order to express $C_k$, we need wide range of
floating numbers, but
in Eqs.~(\ref{Eq:recursive}) we do not need very high
precision.
In other words, we need wide range of the exponent, but
we do not need very large significant numbers.  

We express each real and imaginary parts of $C_k$ in a form of
\be
a \times b^L
\ee
where
\be
1 \le |a| < b
\ee
and $a$ is a double precision real and $L$ is an integer.
When we solve the recursion relation Eqs.~(\ref{Eq:recursive}), we
express all $C_k$, $C_k'$ and $B_k$ in this form. 
The base $b$ can be any number, and we set it to be 8.

To see if this simple trick works or not, we calculate several
cases by this method, and by a high accuracy library, FMLIB\cite{Web:FMLIB}.
We got the same results.
Although this method works for obtaining the coefficients,
$C_i$ in a sufficient double precision, we found a
peculiar configuration on which a huge cancellation
occurs in the sum of $C_i \times \exp(i \mu/T)$ and
the double precision is not enough to get a correct
value of the determinant.

\section{Alternative approach}
In this appendix, we give another possible transformation of the
Wilson fermion determinant. It is more direct extension 
of the Gibbs's approach for the staggered fermion, and may give a general
base.  Unfortunately, in present-days numerical algorithms we cannot
find a reliable one to solve a generalized eigenvalue problem
if involved matrices are singular.  But if in future this problem
is solved, the following can be another good starting point. 

Keeping in mind that for Wilson fermions, 
$(-\kappa(r-\gamma_4)V)^{-1}$ does not exist unless the Wilson term
$r\ne 1$,
we can apply similar transformation in the staggered fermion case by Gibbs 
to the Wilson fermion, and get
\bea
\det \Delta &=& \det \frac{1}{z} \left( 
zB + z^2 (-\kappa(r+\gamma_4)V^\dagger) + (-\kappa(r-\gamma_4)V) \right) \nn
\\
&=&
\det \frac{1}{z} \left( 
zBV + z^2 (-\kappa(r+\gamma_4)) + (-\kappa(r-\gamma_4)V^2) \right)
V^{-1} \nn
\\
&=&
z^{-N} 
\left|
\begin{array}{cccc}
 & & &
 \\
-BV-z(-\kappa(r+\gamma_4)) & & I &
\\
 & & &
\\
  \kappa(r-\gamma_4)V^2       & & -z &
\\
 & & &
\end{array}
\right|
/\det V \nn
\\
&=&
z^{-N} 
\left|
\left(
\begin{array}{cc}
-BV         & I
\\
\kappa(r-\gamma_4)V^2    & 0
\end{array}
\right)
- z
\left(
\begin{array}{cc}
-\kappa(r+\gamma_4)  &  0
\\
  0      &   I
\end{array}
\right)
\right|
\eea

Here the block-matrices are given by 
\bea
&& B V
=
\nonumber
\\
&&
\hspace{-4mm}
\left(
\begin{array}{c|c|ccc|c}
   0 &  B_1V_{12} & 0 & \cdots& &  0 
\\ \hline
   0  & 0 &  B_2V_{23}  &\cdots  &  &  0 
\\ \hline
  0 & 0 & 0 & \cdots & & 
\\ 
 \cdots  &\cdots  &\cdots & \cdots& \cdots  &\cdots
\\ 
   &  & &\cdots    & B_{N_t-2}V_{N_t-2 N_t-1} & 0 
\\ \hline
    0    & 0 & &  \cdots & 0 &   B_{N_t-1}V_{N_t-1 N_t}
\\ \hline
 B_{N_t}V_{N_t 1} & 0 &  &\cdots &0&  0 \\
\end{array}
\right), 
\nonumber
\eea
and 
\bea
&& V^2
=
\nonumber
\\
&&
\hspace{-4mm}
\left(
\begin{array}{c|c|c|cc|c|c}
   0 &  0 & V_{12}V_{23} &  0 & \cdots& &  0 
\\ \hline
   0  & 0 &  0 & V_{23}V_{34}  &\cdots   & &  0 
\\ \hline
  0 & 0 & 0 & \cdots & & &
\\ 
 \cdots  &\cdots  &\cdots & \cdots& \cdots  &\cdots
\\ 
   &  & &\cdots    & & 0 & V_{N_t-2 N_t-1}  V_{N_t-1 N_t}  
\\ \hline
    V_{N_t-1 N_t}  V_{N_t 1}    & 0 & &  \cdots & & 0 &  0
\\ \hline
0 &  V_{N_t 1}V_{12} & 0 &  &\cdots& 0 &  0 \\
\end{array}
\right). 
\nonumber
\eea

By exchange columns and raws,
\bea
&&\left|
\begin{array}{cc}
-BV         & I
\\
\kappa(r-\gamma_4)V^2    & 0
\end{array}
\right|
\rightarrow
 \\
\hspace{-4mm}
&&\left|
\begin{array}{c c|c c|cc|c|cc}
   0 & 0 &  -B_1V_{12} & 1 & \cdots & \cdots& &  0 & 0 
\\ 
   0  & 0 &  \alpha V_{N_1 1}V_{12} & 0 &\cdots & \cdots &  &  0  & 0 
\\ \hline
  0 & 0 & 0 & 0&  -B_2V_{23}   & 1 & \cdots&  &
\\ 
  0 & 0 & 0 & 0&  \alpha V_{12}V_{23}   & 0 & \cdots & & 
\\ \hline
 \cdots & \cdots  &\cdots  &\cdots & \cdots& \cdots & \cdots & & 
\\ 
  \cdots & &  & &\cdots    &  & \cdots &  &
  \\ \hline
 \cdots & \cdots  &\cdots  &\cdots & \cdots& \cdots & \cdots & -B_{N_t-1}V_{N_t-1 N_t} & 1
\\ 
  \cdots & &  & &\cdots    &  & \cdots & \alpha V_{N_t-2 N_t-1}V_{N_t-1 N_t} & 0
\\ \hline
  -B_{N_t}V_{N_t 1}&   1    & 0 & 0 &  \cdots & \cdots &  & 0 & 0   
\\ 
  \alpha V_{N_t-1 N_t}V_{N_t 1} & 0 & 0 & 0 &\cdots & & \cdots &  0 & 0 \\
\end{array}
\right| 
\nn
\eea
Here we introduce $\alpha \equiv \kappa(r-\gamma_4)$.
Applying the same exchange of the columns and raws,
\bea
&&
\left|
\begin{array}{cc}
-\kappa(r+\gamma_4)  &  0
\\
  0      &   I
\end{array}
\right|
\rightarrow
 \\
\hspace{-4mm}
&&\left|
\begin{array}{c c|c c|cc|c|cc}
   -\kappa(r+\gamma_4)  & 0 & 0  & 0 & \cdots & \cdots& &  0 & 0 
\\ 
   0  & 1 & 0  & 0 &\cdots & \cdots &  &  0  & 0 
\\ \hline
  0 & 0 & -\kappa(r+\gamma_4)  & 0&     &  & \cdots&  &
\\ 
  0 & 0 & 0             & 1&     &  & \cdots & & 
\\ \hline
 \cdots & \cdots  &\cdots  &\cdots & \cdots& \cdots & \cdots & & 
\\ 
  \cdots & &  & &\cdots    &  & \cdots &  &
\\ \hline
 0 & 0      &  &  &  \cdots & \cdots &                 & -\kappa(r+\gamma_4)  &   0 
\\ 
 0 & 0     &  &   &  \cdots &             & \cdots     &          0      &  1 \\
\end{array}
\right| 
\nn 
\eea
Then we can write
\be
\det\Delta = z^{-N} \det(T-zS)
\label{Eq:GenEigenVal}
\ee
where
\bea
T
=
\left(
\begin{array}{c|c|ccc|c}
   0 &  t_1 & 0 & \cdots& &  0 
\\ \hline
   0  & 0 & t_2  &\cdots  &  &  0 
\\ \hline
  0 & 0 & 0 & \cdots & & 
\\ 
 \cdots  &\cdots  &\cdots & \cdots& \cdots  &\cdots
\\ 
   &  & &\cdots    & t_{N_t-2} & 0 
\\ \hline
    0    & 0 & &  \cdots & 0 &  t_{N_t-1} 
\\ \hline
t_{N_t} & 0 &  &\cdots &0&  0 \\
\end{array}
\right) ,
\nonumber
\eea

\be
t_i = 
\matThree
-B_i V_{i,i+1}                        &  & 1
\\
\kappa (r-\gamma_4)V_{i-1,i}V_{i,i+1} &  & 0
\emat
\ee

and
\be
S = 
\left(
\begin{array}{c|c|ccc|c}
   s & 0  & 0 & \cdots& &  0 
\\ \hline
   0  & s & 0  &\cdots  &  &  0 
\\ \hline
  0 & 0 & 0 & \cdots & & 
\\ 
 \cdots  &\cdots  &\cdots & \cdots& \cdots  &\cdots
\\ \hline
    0    & 0 & &  \cdots & s &  0 
\\ \hline
   0 & 0 &  &\cdots &0&  s \\
\end{array}
\right). 
\nonumber
\ee

\be
s = 
\left(
\begin{array}{cc}
-\kappa(r+\gamma_4)  &  0
\\
  0      &   I
\end{array}
\right) .
\ee
Each $t_i$ and $s$ is $(4N_cN_xN_yN_z)\times(4N_cN_xN_yN_z)$
matrix.

Eq.~(\ref{Eq:GenEigenVal}) is a form of the generalized
eigenvalue problem\cite{GenEigenVal}.
There is a mathematical theorem (Generalized Schur Decomposition)
which tells us that there exist unitary matrices $Y$ and $Z$ such that
$Y^{\dagger} S Z$ and $Y^{\dagger} T Z$ are upper triangular.
Let $\alpha_k$ and $\beta_k$ be diagonal elements of
these matrices. Then
\be
\det(T-zS) = \det Y Z^\dagger \prod_k (\alpha_k - z \beta_k)
\ee
Half of $\alpha$'s and $\beta$'s vanish.

This formula has an advantage that $T$ and $S$ do not 
have inverse matrix like $Q$ in Eq.~(\ref{May1010eq2}),
and can be easily constructed.
A problem is that matrices $T$ and $S$ are singular.  
To our knowledge, no stable algorithm is know to solve the 
generalized eigenvalue problem in such a case.

\end{document}